\documentclass[11pt,a4paper]{article}
\pdfoutput=1
\usepackage{jheppub}
\usepackage[utf8]{inputenc}
\usepackage{mathpazo, color, hyperref, dsfont, slashed, multirow, amssymb, amsmath, amsfonts, comment, float, latexsym, graphicx, epstopdf, float, fancyvrb, chngcntr, etoolbox, booktabs, bbm, epsfig, soul, pdfpages, array, lipsum, amssymb, graphicx, hepunits, slashed, ccaption, pdfpages, scrextend, makecell, cellspace, adjustbox}
\usepackage[titletoc, toc, title]{appendix}

\hypersetup{bookmarks=true, unicode=true, pdftoolbar=true, pdfmenubar=true,
 pdffitwindow=false, pdfstartview={FitH}, pdftitle={},
 pdfauthor={}, pdfsubject={Subject},
 pdfcreator={Creator}, pdfproducer={Producer}, pdfkeywords={}{Beyond the Standard Model Physics},
 pdfnewwindow=true, colorlinks=true,
 linkcolor=blue, citecolor=magenta, filecolor=magenta, urlcolor=cyan}
\title{A Scotogenic Model with Two Inert Doublets}

\author{Amine Ahriche $^{\href{https://orcid.org/0000-0003-0230-1774}{\includegraphics[width=2.5mm]{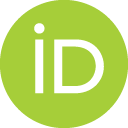}}}$}
\emailAdd{ahriche@sharjah.ac.ae}

\affiliation{Department of Applied Physics and Astronomy, University of Sharjah, P.O. Box 27272 Sharjah, UAE.}

\abstract{In this work, we present a scotogenic model, where the neutrino mass is generated at one-loop diagrams. The standard model (SM) is extended by three singlet Majorana fermions and two inert scalar doublets instead of one doublet as in the minimal scotogenic model. The model scalar sector includes two CP-even, two CP-odd and two charged scalars in addition to the Higgs.
The dark matter (DM) candidate could be either the light Majorana fermion (Majorana DM), or the lightest among the CP-even and the CP-odd scalars (scalar DM). We show that the model accommodates both Majorana and scalar DM within a significant viable parameter space, while considering all the relevant theoretical and experimental constraints such as perturbativity, vacuum stability, unitarity, the di-photon Higgs decay, electroweak precision tests and lepton flavor violating constraints. In addition to the collider signatures predicted by the minimal scotogenic model, our model predicts some novel signatures that can be probed through some final states such as $8~jets+\slashed{E}_T$, $1\ell+4~jets+\slashed{E}_T$ and $4b+\slashed{E}_T$.}

\begin{document}

\maketitle

\section{Introduction \label{sec:introduction}}

The standard model (SM) of particle physics was successful in describing
the properties of elementary particles and their interactions around
the electroweak (EW) scale. However, there are still unanswered questions
such as the neutrino oscillation data~and the dark matter (DM) nature.
Neutrino oscillation data established that at least two of the three
SM neutrinos have mass and nonzero mixing. However, their properties,
such as their nature and the origin of the smallness of their mass,
have no explanation within the SM, which demands for new physics.
One of the most popular mechanisms to explain why neutrino mass is
tiny; is the so-called seesaw mechanism~\cite{Mohapatra:1986aw},
which assumes the existence of right-handed (RH) neutrinos with masses
that are heavier by several orders of magnitude than the EW scale.
Such heavy particles decouple from the low-energy spectrum at the
EW scale, and hence, cannot be directly probed at high-energy physics
experiments. Another approach that requires lower new physics scale,
where small neutrino mass is generated naturally; is via quantum corrections,
where loop suppression factor(s) ensures the neutrino mass smallness.
This can be realized at one loop~\cite{Zee:1985rj}, two loops~\cite{Zee-86},
three loops~\cite{Aoki:2008av,Krauss:2002px}, or four loops~\cite{Nomura:2016fzs}.
Interesting features for this TeV scale new physics class of models
can be tested at high energy colliders~\cite{Cheung:2004xm,Ahriche:2014xra}
(For a review, see~\cite{Cai:2017jrq}).

The simplest and most popular realization of a radiative neutrino
mass mechanism is the so-called scotogenic model~\cite{Ma:2006km},
where the SM is extended by an inert scalar doublet and three singlet
Majorana fermions. In addition, this model could accommodate two possible
DM candidates, i.e., the lightest among the Majorana fermions and
the light neutral scalar in the inert dark doublet. The scalar DM
scenario is identical to the inert doublet Higgs model (IDM), that
is highly constrained by the DM direct detection experiments. In the
scenario of Majorana DM, the right amount of the DM relic density
requires relatively large values for the new Yukawa couplings that
couple the inert doublet with the Majorana fermions~\cite{Ahriche:2017iar,Ahriche:2018ger}.
This can be achieved only by imposing a strong degeneracy between
the masses of the CP-even and CP-odd scalars, making the quartic coupling
suppressed $\lambda_{5}\sim10^{-10}-10^{-9}$. Such fine-tuning can
be avoided by extending the minimal scotogenic model (MSctM) by two
real singlets and impose a global $Z_{4}/Z_{2}$ symmetry~\cite{Ahriche:2020pwq}.
Indeed, there have many models that have been proposed and studied
beyond the MSctM~\cite{Escribano:2020iqq,Ahriche:2016cio}. For example,
in~\cite{Escribano:2020iqq} the authors studied a generalized scotogenic
model, or what they called "\textit{general scotogenic model}",
by considering $n_{N}$ Majorana singlet fermions and $n_{\eta}$
inert doublets, where many phenomenological aspects were discussed.

Based on the current/future results of both ATLAS and CMS, the IDM
(and therefore, the MSctM with scalar DM model) parameter space will
become more constrained as the precision measurements of the EW sector
gets improved. In addition, the recent results from the direct detection
experiments such as LUX-ZEPLIN experiment~\cite{LUX-ZEPLIN:2022qhg},
would place more stringent bounds on the model. This makes the extension
of the IDM interesting, that might evade some of the constraints and
predict additional interesting signatures. On top of that, among the
important motivations to extend the IDM; is that it does not allow
CP violation in the scalar sector, where an additional $SU(2)_{L}$
Higgs doublet is useful to do this task~\cite{Grzadkowski:2009bt}.
According to the previously mentioned motivations, it is natural to
consider the realization of a scotogenic model where the SM is extended
by three singlet Majorana fermions $N_{1,2,3}$ and two inert doublets
$\Phi_{1,2}$ that couple to the leptonic left-handed doublets and
the singlet Majorana's. Here, the new additional fields transform
as odd fields under a global $Z_{2}$ symmetry $\{N_{1,2,3},\Phi_{1,2}\}\rightarrow\{-N_{1,2,3},-\Phi_{1,2}\}$
to ensure the DM stability.

This work is organized as follow. In section~\ref{sec:model}, we
present the model and show the neutrino mass generation at loop. In
section~\ref{sec:constraints}, different theoretical and experimental
constraints are discussed. Sections~\ref{sec:NumAn} and~\ref{sec:DM}
are devoted to show the viable parameter space and the DM phenomenology,
respectively, in both scenarios of Majorana and scalar DM candidates.
We conclude in section~\ref{sec:conclusion}. In addition, the paper
includes three appendices: in Appendix~\ref{app1}, we discuss how
to estimate the new Yukawa couplings where a generalized parameterization
like the Casas-Ibarra one is introduced. In Appendix~\ref{sec:nInert},
we generalize our results for the case of a scotogenic model with
\textbf{n} inert scalar doublets instead of two. In Appendix~\ref{Uni},
we give the amplitude matrices that are relevant to the perturbative
unitarity conditions.

\section{The Model \& Neutrino Mass\label{sec:model}}

Here, we extend the SM by two inert doublets denoted by $\Phi_{1,2}\sim(1,2,1)$
and three singlet Majorana fermions $N_{i}\sim(1,1,0)$. The Lagrangian
that involves the Majorana fermions can be written as
\begin{eqnarray}
\mathcal{L} & \supset & \overline{L}_{\alpha}\big(h_{\alpha,k}~\epsilon.\Phi_{1}+h_{\alpha,k+3}~\epsilon.\Phi_{2}\big)N_{k}+\frac{1}{2}\bar{N}_{k}^{C}M_{k}N_{k}+h.c.,\label{LL}
\end{eqnarray}
where $L_{\alpha}$ are the left-handed lepton doublets, and $\epsilon=i\sigma_{2}$
is an anti-symmetric tensor. Here, $h_{\alpha,k}$ is a $3\times6$
new Yukawa matrix, the mass matrix $M_{k}$ can be considered diagonal.

The most general $Z_{2}$-symmetric, renormalizable, and gauge invariant
potential reads
\begin{eqnarray}
V\left(\mathcal{H},\Phi,S,\chi\right) & = & -\mu_{H}^{2}|\mathcal{H}|^{2}+\mu_{i}^{2}|\Phi_{i}|^{2}+\frac{\lambda_{H}}{6}|\mathcal{H}|^{4}+\frac{\lambda_{i}}{6}|\Phi_{i}|^{4}+\omega_{i}|\mathcal{H}|^{2}|\Phi_{i}|^{2}+\kappa_{i}|\mathcal{H}^{\dagger}\Phi_{i}|^{2}\nonumber \\
 &  & +\varrho_{1}|\Phi_{1}|^{2}|\Phi_{2}|^{2}+\varrho_{2}|\Phi_{1}^{\dagger}\Phi_{2}|^{2}+\left\{ \mu_{3}^{2}\Phi_{1}^{\dagger}\Phi_{2}+\frac{1}{2}\xi_{i}(\mathcal{H}^{\dagger}\Phi_{i})^{2}+\xi_{3}(\mathcal{H}^{\dagger}\Phi_{1})(\mathcal{H}^{\dagger}\Phi_{2})\right.\nonumber \\
 &  & \left.+\xi_{4}(\Phi_{1}^{\dagger}\mathcal{H})(\mathcal{H}^{\dagger}\Phi_{2})+h.c.\right\} ,\label{eq:V}
\end{eqnarray}
with $\mathcal{H}$, and $\Phi$ can be parameterized as follows
\begin{equation}
\mathcal{H}=\left(\begin{array}{c}
\chi^{+}\\
\frac{1}{\sqrt{2}}(\upsilon+h+i\chi^{0})
\end{array}\right),~\Phi_{i}=\left(\begin{array}{c}
S_{i}^{+}\\
\frac{1}{\sqrt{2}}(S_{i}^{0}+iQ_{i}^{0})
\end{array}\right).\label{eq:para}
\end{equation}

The terms of the Lagrangian in (\ref{LL}) and (\ref{eq:V}) are invariant
under a global $Z_{2}$ symmetry according to the charges
\begin{equation}
\Phi_{1,2}\rightarrow-\Phi_{1,2},\,N_{1,2,3}\rightarrow-N_{1,2,3},
\end{equation}
where all other fields are even. This $Z_{2}$ symmetry forbids such
terms like $(\mathcal{H}^{\dagger}\Phi_{i})(\Phi_{j}^{\dagger}\Phi_{k})$.
In case of complex values for $\mu_{3}^{2}$ and/or $\xi_{i}$, CP
is explicitly broken, where we obtain CP-violating interactions relevant
to the four neutral inert eigenstates in the basis \{$S_{1}^{0},S_{2}^{0}$,$Q_{1}^{0},Q_{2}^{0}$\}.
This case would be very interesting since the additional CP-violating
sources are required for matter antimatter asymmetry generation~\cite{Sarma:2020msa}.
In what follows, we consider real values for the parameters $\mu_{3}^{2}$
and $\xi_{i=1,2,3,4}$ to avoid CP violation. The case of explicit
CP violation (complex values for $\mu_{3}^{2}$ and/or $\xi_{i}$)
requires an independent investigation since it may have important
cosmological consequences~.

After the electroweak symmetry breaking (EWSB), we are left with three
$CP$-even scalars $(h,H_{1,2}^{0})$, two $CP$-odd scalar $A_{1,2}^{0}$
and two pair of charged scalars $H_{1,2}^{\pm}$. The Higgs mass is
given $m_{h}^{2}=2\mu_{H}^{2}=\frac{\lambda_{H}}{3}\upsilon^{2}$,
and the inert eigenstates are defined as
\begin{align}
\left(\begin{array}{c}
H_{1}^{0}\\
H_{2}^{0}
\end{array}\right) & =\left(\begin{array}{cc}
c_{H} & s_{H}\\
-s_{H} & c_{H}
\end{array}\right)\left(\begin{array}{c}
S_{1}^{0}\\
S_{2}^{0}
\end{array}\right),\left(\begin{array}{c}
A_{1}^{0}\\
A_{2}^{0}
\end{array}\right)=\left(\begin{array}{cc}
c_{A} & s_{A}\\
-s_{A} & c_{A}
\end{array}\right)\left(\begin{array}{c}
Q_{1}^{0}\\
Q_{2}^{0}
\end{array}\right),\nonumber \\
 & \left(\begin{array}{c}
H_{1}^{\pm}\\
H_{2}^{\pm}
\end{array}\right)=\left(\begin{array}{cc}
c_{C} & s_{C}\\
-s_{C} & c_{C}
\end{array}\right)\left(\begin{array}{c}
S_{1}^{\pm}\\
S_{2}^{\pm}
\end{array}\right),\label{eq:EigS}
\end{align}
with $c_{X}=\cos\theta_{X},\,s_{X}=\sin\theta_{X}$ and $\theta_{H}$
, $\theta_{A}$ and $\theta_{C}$ are the mixing angles that diagonalise
the $CP$-even, $CP$-odd and charged scalars mass matrices, respectively.
The charged, neutral $CP$-even and $CP$-odd scalar squared mass
matrices in the basis \{$S_{1}^{\pm},S_{2}^{\pm}$\}, \{$S_{1}^{0},S_{2}^{0}$\}
and \{$Q_{1}^{0},Q_{2}^{0}$\}, respectively, are given by
\begin{align}
M_{C}^{2} & =\left[\begin{array}{cc}
\mu_{1}^{2} & \mu_{3}^{2}\\
\mu_{3}^{2} & \mu_{2}^{2}
\end{array}\right]+\frac{\upsilon^{2}}{2}\left[\begin{array}{cc}
\omega_{1} & 0\\
0 & \omega_{2}
\end{array}\right],\,M_{H,A}^{2}=M_{C}^{2}+\frac{\upsilon^{2}}{2}\left[\begin{array}{cc}
\kappa_{1} & \xi_{4}\\
\xi_{4} & \kappa_{2}
\end{array}\right]\pm\frac{\upsilon^{2}}{2}\left[\begin{array}{cc}
\xi_{1} & \xi_{3}\\
\xi_{3} & \xi_{2}
\end{array}\right].
\end{align}

The mass eigenvalues and the mixing are given by
\begin{align}
m_{X_{1,2}}^{2} & =\frac{1}{2}\bigg([M_{X}^{2}]_{11}+[M_{X}^{2}]_{22}\mp\sqrt{([M_{X}^{2}]_{22}-[M_{X}^{2}]_{11})^{2}+4[M_{X}^{2}]_{12}^{2}}\bigg),\nonumber \\
\tan(2\theta_{X}) & =2[M_{X}^{2}]_{12}/\Big([M_{X}^{2}]_{22}-[M_{X}^{2}]_{11}\Big),
\end{align}
for ($[X_{1,2},\theta_{X}]=[H_{1,2}^{0},\theta_{H}],[A_{1,2}^{0},\theta_{A}],[H_{1,2}^{\pm},\theta_{C}]$),
respectively.

Then, the independent free parameters are
\begin{equation}
M_{1,2,3},~m_{H_{1,2}^{\pm}},~m_{H_{1,2}^{0}},~m_{A_{1,2}^{0}},~s_{H,A,C},~\omega_{1,2},~h_{\alpha i},\label{eq:parfree}
\end{equation}
where
\begin{align}
\mu_{1}^{2} & =m_{H_{1}^{\pm}}^{2}c_{C}^{2}+m_{H_{2}^{\pm}}^{2}s_{C}^{2}-\frac{1}{2}\omega_{1}\upsilon^{2},\,\mu_{2}^{2}=m_{H_{1}^{\pm}}^{2}s_{C}^{2}+m_{H_{2}^{\pm}}^{2}c_{C}^{2}-\frac{1}{2}\omega_{2}\upsilon^{2},\,\mu_{3}^{2}=c_{C}s_{C}(m_{H_{2}^{\pm}}^{2}-m_{H_{1}^{\pm}}^{2}),\nonumber \\
\kappa_{1}\upsilon^{2} & =m_{H_{1}^{0}}^{2}c_{H}^{2}+m_{H_{2}^{0}}^{2}s_{H}^{2}+m_{A_{1}^{0}}^{2}c_{A}^{2}+m_{A_{2}^{0}}^{2}s_{A}^{2}-2(m_{H_{1}^{\pm}}^{2}c_{C}^{2}+m_{H_{2}^{\pm}}^{2}s_{C}^{2}),\nonumber \\
\kappa_{2}\upsilon^{2} & =m_{H_{2}^{0}}^{2}c_{H}^{2}+m_{H_{1}^{0}}^{2}s_{H}^{2}+m_{A_{2}^{0}}^{2}c_{A}^{2}+m_{A_{1}^{0}}^{2}s_{A}^{2}-2(m_{H_{2}^{\pm}}^{2}c_{C}^{2}+m_{H_{1}^{\pm}}^{2}s_{C}^{2}),\nonumber \\
\xi_{1}\upsilon^{2} & =(m_{H_{1}^{0}}^{2}c_{H}^{2}+m_{H_{2}^{0}}^{2}s_{H}^{2})-(m_{A_{2}^{0}}^{2}s_{A}^{2}+m_{A_{1}^{0}}^{2}c_{A}^{2}),~\xi_{2}\upsilon^{2}=(m_{H_{1}^{0}}^{2}s_{H}^{2}+m_{H_{2}^{0}}^{2}c_{H}^{2})-(m_{A_{1}^{0}}^{2}s_{A}^{2}+m_{A_{2}^{0}}^{2}c_{A}^{2}),\nonumber \\
\xi_{3}\upsilon^{2} & =s_{H}c_{H}(m_{H_{2}^{0}}^{2}-m_{H_{1}^{0}}^{2})-s_{A}c_{A}(m_{A_{2}^{0}}^{2}-m_{A_{1}^{0}}^{2}),\nonumber \\
\xi_{4}\upsilon^{2} & =s_{H}c_{H}(m_{H_{2}^{0}}^{2}-m_{H_{1}^{0}}^{2})+s_{A}c_{A}(m_{A_{2}^{0}}^{2}-m_{A_{1}^{0}}^{2})-2c_{C}s_{C}(m_{H_{2}^{\pm}}^{2}-m_{H_{1}^{\pm}}^{2}).
\end{align}

The first term in (\ref{LL}), can be written as
\begin{align}
\mathcal{L} & \supset\sum_{\alpha,j,k}\frac{1}{\sqrt{2}}\big(g_{\alpha k}^{(j)}H_{j}^{0}+if_{\alpha k}^{(j)}A_{j}^{0}\big)\overline{\nu}_{\alpha L}N_{k}-\zeta_{\alpha k}^{(j)}H_{j}^{\pm}\overline{\ell}_{\alpha L}N_{k}+h.c,\label{Lfl}
\end{align}
with $\alpha=e,\mu,\tau$, $j=1,2$ and $k=1,3$; and
\begin{align}
g_{\alpha k}^{(1)} & =c_{H}h_{\alpha,k}+s_{H}h_{\alpha,k+3},\,\,g_{\alpha k}^{(2)}=-s_{H}h_{\alpha,k}+c_{H}h_{\alpha,k+3},\nonumber \\
f_{\alpha k}^{(1)} & =c_{A}h_{\alpha,k}+s_{A}h_{\alpha,k+3},\,\,f_{\alpha k}^{(2)}=-s_{A}h_{\alpha,k}+c_{A}h_{\alpha,k+3},\nonumber \\
\zeta_{\alpha k}^{(1)} & =c_{C}h_{\alpha,k}+s_{C}h_{\alpha,k+3},\,\,\zeta_{\alpha k}^{(2)}=-s_{C}h_{\alpha,k}+c_{C}h_{\alpha,k+3}.\label{coup}
\end{align}

\begin{figure}[h]
\begin{centering}
\includegraphics[width=0.48\textwidth]{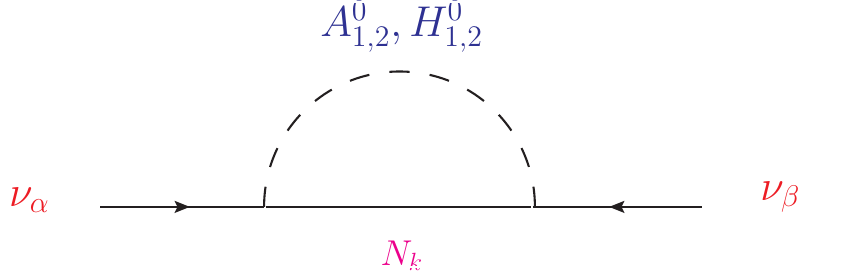}
\par\end{centering}
\caption{The 12 Feynman diagrams responsible for neutrino mass generation.}
\label{fig:ms}
\end{figure}

The first two interaction terms in (\ref{Lfl}) give rise to the neutrino
mass a la scotogenic way; and the last term leads to the LFV processes
$\ell_{\beta}\rightarrow\ell_{\alpha}+\gamma$ and $\ell_{\beta}\rightarrow3\ell_{\alpha}$.
Then, the neutrino mass $[\alpha,\beta]$ matrix element comes from
the contribution 12 diagrams as shown in Fig.~\ref{fig:ms}.

It can be written as
\begin{align}
m_{\alpha\beta}^{(\nu)} & =\sum_{k=1}^{3}\sum_{j=1}^{2}M_{k}\Big\{ g_{\alpha k}^{(j)}\,g_{\beta k}^{(j)}\,\mathcal{F}\Big(m_{H_{j}^{0}}/M_{k}\Big)-f_{\alpha k}^{(j)}\,f_{\beta k}^{(j)}\,\mathcal{F}\Big(m_{A_{j}^{0}}/M_{k}\Big)\Big\},\label{eq:mnu}
\end{align}
with $\mathcal{F}(x)=\frac{x^{2}\log(x)}{8\pi^{2}(x^{2}-1)}$. The
divergent parts of the diagrams in Fig.~\ref{fig:ms} that are mediated
by $H_{j}^{0}$ and $A_{j}^{0}$ cancels each other due to the identity\footnote{One has to mention that this identity is equivalent to (18) in~\cite{Escribano:2020iqq},
where the scotogenic model is generalized by considering many Majorana
singlets and inert doublets.}
\begin{equation}
\sum_{j=1}^{2}\Big\{ g_{\alpha k}^{(j)}\,g_{\beta k}^{(j)}-f_{\alpha k}^{(j)}\,f_{\beta k}^{(j)}\Big\}=0.
\end{equation}

The neutrino mass matrix (\ref{eq:mnu}) can be parameterized as
\begin{align}
(m^{(\nu)})_{3\times3} & =(h)_{3\times6}.(\Lambda)_{6\times6}.(h^{T})_{6\times3}\,\,\,\,,\label{eq:CIb}\\
\Lambda & =\left(\begin{array}{cc}
\varpi_{1} & \varpi_{3}\\
\varpi_{3} & \varpi_{2}
\end{array}\right),
\end{align}
where $\varpi_{1,2,3}$ are $3\times3$ diagonal matrices whose elements
are given by
\begin{align}
(\varpi_{1})_{ik} & =\delta_{ik}M_{k}\big[c_{H}^{2}\mathcal{F}\Big(m_{H_{1}^{0}}/M_{k}\Big)+s_{H}^{2}\,\,\mathcal{F}\Big(m_{H_{2}^{0}}/M_{k}\Big)-c_{A}^{2}\,\mathcal{F}\Big(m_{A_{1}^{0}}/M_{k}\Big)-s_{A}^{2}\,\,\mathcal{F}\Big(m_{A_{2}^{0}}/M_{k}\Big)\big],\nonumber \\
(\varpi_{2})_{ik} & =\delta_{ik}M_{k}\big[s_{H}^{2}\mathcal{F}\Big(m_{H_{1}^{0}}/M_{k}\Big)+c_{H}^{2}\,\,\mathcal{F}\Big(m_{H_{2}^{0}}/M_{k}\Big)-s_{A}^{2}\,\mathcal{F}\Big(m_{A_{1}^{0}}/M_{k}\Big)-c_{A}^{2}\,\mathcal{F}\Big(m_{A_{2}^{0}}/M_{k}\Big)\big],\nonumber \\
(\varpi_{3})_{ik} & =\delta_{ik}M_{k}\big[s_{H}c_{H}(\mathcal{F}\Big(m_{H_{1}^{0}}/M_{k}\Big)-\,\mathcal{F}\Big(m_{H_{2}^{0}}/M_{k}\Big))-s_{A}c_{A}(\,\mathcal{F}\Big(m_{A_{1}^{0}}/M_{k}\Big)-\,\mathcal{F}\Big(m_{A_{2}^{0}}/M_{k}\Big))\big],
\end{align}
for $i,k=1,3$. Here, the matrix $\Lambda$ in (\ref{eq:CIb}) is
not diagonal, and therefore can cannot use the Casas-Ibarra parameterization~\cite{Casas:2001sr}
to estimate the Yukawa couplings $h$. However, we can derive a similar
formula in such case ($\Lambda$ is not diagonal). We give a general
solution for the Yukawa coupling $3\times M$ matrix\footnote{Here, $M=6$ in our case. However, the solutions (\ref{eq:h}) are
valid for any value $M\geq3$.} as
\begin{equation}
h_{3\times M}=(U_{\nu})_{3\times3}.(D_{\sqrt{m_{\nu}}})_{3\times3}.(T)_{3\times M}.(D_{(\Lambda'_{i})^{-1/2}})_{M\times M}.(Q)_{M\times M}\,\,\,\,\,\,,\label{eq:h}
\end{equation}
with $U_{\nu}$ is the Pontecorvo-Maki-Nakawaga-Sakata (PMNS) mixing
matrix, $D_{\sqrt{m_{\nu}}}=\textrm{diag}\{m_{1}^{1/2}$ $,m_{2}^{1/2},m_{3}^{1/2}\}$
with $m_{i}$ are the neutrino mass eigenvalues; and $T$ is an arbitrary
$3\times M$ matrix that fulfills the identity $T_{3\times M}.T_{M\times3}^{T}=\boldsymbol{1}_{3\times3}$.
The matrix $(D_{(\Lambda'_{i})^{-1/2}})_{M\times M}=\textrm{diag}\left\{ (\Lambda'_{i})^{-1/2}\right\} $,
with $\Lambda'_{i}$ to be the eigenvalues of the matrix $\Lambda$
in (\ref{eq:CIb}), that diagonalized using the matrix $Q$, i.e.,
$\textrm{diag}\left\{ \Lambda'_{i}\right\} =Q.\Lambda.Q^{T}$. In
Appendix \ref{app1}, we discuss the derivation of (\ref{eq:h}) and
the possible representations for this matrix. In addition, we discuss
in Appendix \ref{sec:nInert} the generalization of this model by
considering \textbf{n} inert scalar doublets instead of two doublets.

\section{Theoretical \& Experimental Constraints~\label{sec:constraints}}

Here, we discuss the theoretical and the experimental constraints
relevant to our model. These constraints are briefly discussed below:
\begin{itemize}
\item \textbf{Perturbativity}: all the quartic vertices of the scalar fields
should be satisfy the perturbativity bounds, i.e.,
\begin{align}
 & \max\big\{\lambda_{H},\lambda_{1,2},|\omega_{1,2}|,|\kappa_{1,2}+\omega_{1,2}|,|\omega_{1,2}+\kappa_{1,2}\pm\xi_{1,2}|,|\varrho_{1}|,|\varrho_{1}+\varrho_{2}|,\nonumber\\
 & |\xi_{1,2}|,\frac{1}{2}|\kappa_{1,2}\pm\xi_{1,2}|,\frac{1}{2}|\varrho_{2}|\big\}\leq4\pi.
\end{align}

\item \textbf{Perturbative unitarity}: the perturbative unitarity has to
be preserved in all processes involving scalars or gauge bosons. In
the high-energy limit, the gauge bosons should be replaced by their
Goldstones; and the scattering amplitude matrix of $\phi_{i}~\phi_{j}\rightarrow\phi_{k}\phi_{m}$
can be easily calculated. It has been shown that the perturbative
unitarity conditions are achieved if the eigenvalues of the scattering
amplitude matrix to be smaller than $|\Lambda_{i}|<8\pi$~\cite{Akeroyd:2000wc}.

In our model, due to some exact symmetries such as the electric charge,
CP and the global $Z_{2}$ symmetry, the full scattering amplitude
matrix can be divided into six sub-matrices. These six sub-matrices
are defined in the basis of initial/final states that are: (1) CP-even,
$Z_{2}$ even, \& $Q_{em}=0$; (2) CP-even, $Z_{2}$ odd, \& $Q_{em}=0$,
(3) CP-odd, $Z_{2}$ even, \& $Q_{em}=0$, (4) CP-odd, $Z_{2}$ odd,
\& $Q_{em}=0$, (5) $Z_{2}$ even, \& $Q_{em}=\pm1$; and (6) $Z_{2}$
odd, \& $Q_{em}=\pm1$. These sub-matrices and their basis are given
in Appendix \ref{Uni} in details.
\item \textbf{Vacuum Stability}: the scalar potential is required to be
bounded from below in all the directions of the field space. It is
obvious that along the pure directions ($\Phi_{1}=\Phi_{2}=0$), ($\mathcal{H}=\Phi_{2}=0$)
and ($\mathcal{H}=\Phi_{1}=0$), the vacuum stability conditions,
i.e., the potential positivity at large field values, are $\lambda_{H},\lambda_{1},\lambda_{2}>0$,
respectively. However, along any direction beside the pure ones, the
potential positivity is guaranteed if all quartic couplings are positive.
In case of negative quartic coupling(s), the vacuum stability conditions
are~\cite{Kannike:2012pe}:
\begin{equation}
\left|\begin{array}{ccc}
\lambda_{H} & \overline{\omega_{1}+\kappa_{1}+\xi_{1}}, & \overline{\omega_{2}+\kappa_{2}+\xi_{2}}\\
\overline{\omega_{1}+\kappa_{1}+\xi_{1}} & \lambda_{1} & \overline{\varrho_{1}+\varrho_{2}}\\
\overline{\omega_{2}+\kappa_{2}+\xi_{2}} & \overline{\varrho_{1}+\varrho_{2}} & \lambda_{2}
\end{array}\right|>0,~~\left|\begin{array}{ccc}
\frac{2}{3}\lambda_{H} & \overline{\omega_{1}+\kappa_{1}}, & \overline{\omega_{2}+\kappa_{2}}\\
\overline{\omega_{1}+\kappa_{1}} & \frac{2}{3}\lambda_{1} & \overline{\varrho_{1}+\varrho_{2}}\\
\overline{\omega_{2}+\kappa_{2}} & \overline{\varrho_{1}+\varrho_{2}} & \frac{2}{3}\lambda_{2}
\end{array}\right|>0,\label{eq:Vac}
\end{equation}
for the neutral and charged fields directions, respectively, with
$\overline{X}=\min\left(X,0\right)$.

In addition, we require that the vacuum $(<h>=\upsilon,<S_{i}^{0}>=0)$
would be the deepest one. As a conservative choice in this work, we
consider the conditions that the potential (\ref{eq:V}) should not
develop a vev in the inert directions, i.e., we should have $<S_{i}^{0}>=0)$
either at $<h>=0$ or $<h>=\upsilon$. These conditions can be translated
into
\begin{equation}
\mu_{1}^{2},\mu_{2}^{2},\mu_{1}^{2}+\mu_{2}^{2}-\sqrt{(\mu_{2}^{2}-\mu_{1}^{2})^{2}+4(\mu_{3}^{2})^{2}}>0,\label{eq:WrVac}
\end{equation}
and obviously, $m_{H_{1}^{0}}^{2},m_{A_{1}^{0}}^{2},m_{H_{1}^{\pm}}^{2}>0$.
\item \textbf{Gauge bosons decay widths}: The decay widths of the $W/Z$-bosons
were measured with high precision at LEP. Therefore, we require that
the decays of $W/Z$-bosons to $Z_{2}$-odd scalars is closed. This
is fulfilled if one assumes that
\begin{equation}
\min\big(m_{H_{1}^{0}}+m_{A_{1}^{0}},\,2m_{H_{1}^{\pm}}\big)>m_{Z},\,\min\big(m_{H_{1}^{\pm}}+m_{A_{1}^{0}},\,m_{H_{1}^{\pm}}+m_{H_{1}^{0}}\big)>m_{W}.
\end{equation}
\item \textbf{Lepton flavor violating (LFV) decays}: in this model, LFV
decay processes arise via one-loop diagrams mediated by the $H_{i=1,2}^{\pm}$
and $N_{k}$ particles. The branching ratio of the decay $\mathcal{B}(\ell_{\alpha}\rightarrow\ell_{\beta}+\gamma)$
due to the contribution of the interactions (\ref{LL}) is~\cite{Toma:2013zsa}:
\begin{align}
\mathcal{B}(\ell_{\alpha} & \rightarrow\ell_{\beta}+\gamma)=\frac{3\alpha\upsilon^{4}}{32\pi}\left|\sum_{k=1}^{3}\sum_{j=1}^{2}\frac{\big(\zeta_{\alpha k}^{(j)}\big)\big(\zeta_{\beta k}^{(j)}\big)^{*}}{m_{H_{j}^{\pm}}^{2}}F\Big(\frac{M_{k}^{2}}{m_{H_{j}^{\pm}}^{2}}\Big)\right|^{2},\label{LFV}
\end{align}
where $\alpha=e^{2}/4\pi$ is the electromagnetic fine structure constant
and $F(x)=(1-6x+3x^{2}+2x^{3}-6x^{2}\log x)/6(1-x)^{4}.$ Since the
bounds on $\ell_{\alpha}\rightarrow\ell_{\beta}+\gamma$ are more
severe, especially $\mu\rightarrow e+\gamma$, we will consider only
the LFV bounds on $\ell_{\alpha}\rightarrow\ell_{\beta}+\gamma$ and
not $\ell_{\alpha}\rightarrow\ell_{\beta}\ell_{\beta}\ell_{\beta}$,
since the latter's would implicitly fulfilled~\cite{Chekkal:2017eka}.

One has to notice that getting experimentally allowed values for the
branching ratios (\ref{LFV}) can be achieved by considering very
small Yukawa couplings $h_{\alpha i}$ and/or very heavy charged scalars.
Although, this choice is preferred by the scalar DM case, but not
interesting from Majorana DM relic density and/or collider point of
view. However, one can consider well chosen values for the Yukawa
couplings ($h_{\alpha i}$) and the Majorana and charged scalar masses,
where the terms in the summation in (\ref{LFV}) cancel each other.
This allows relatively large Yukawa couplings as well relatively light
charged scalars in order to accommodate the DM relic density in the
case of Majorana DM~\cite{Ahriche:2017iar,Ahriche:2020pwq}; and
provide interesting predictions at colliders~\cite{Ahriche:2014xra,Chekkal:2017eka,Ahriche:2017iar,Ahriche:2018ger}.
This can be clearly seen when comparing the Yukawa couplings and the
Majorana and scalar masses for the two benchmark points (BPs) in Table~\ref{BPs}
that corresponds to Majorana and scalar DM cases.
\item \textbf{Direct searches of charginos and neutralinos at the LEP-II
experiment}: we use the null results of neutralinos and charginos
at LEP~\cite{DELPHI:2003uqw} to put lower bounds on the masses of
charged Higgs $H_{1}^{\pm}$ and the light neutral scalars of the
inert doublets ($H_{1}^{0},A_{1}^{0}$). The bound obtained from a
re-interpretation of neutralino searches~\cite{Lundstrom:2008ai}
cannot apply to our model since the decays $A_{1}^{0}\to H_{1}^{\pm}Z(\to\ell^{+}\ell^{-})$
are kinematically forbidden. On the other hand, in most regions of
the parameter space, the charged Higgs decays exclusively into a Majorana
fermion and a charged lepton. For Yukawa couplings of order one $h_{ei}\simeq\mathcal{O}(1)$,
the following bounds are derived $m_{H^{\pm}}>100~\textrm{GeV}$~\cite{Ahriche:2018ger}.
\item \textbf{The electroweak precision tests}: in this model, the oblique
parameters acquire contributions from the existence of inert scalars.
We take $\Delta U=0$ in our analysis, the oblique parameters are
given by~\cite{Grimus:2008nb}
\begin{align}
\varDelta T & =\frac{1}{16\pi s_{\mathrm{w}}^{2}M_{W}^{2}}\left\{ c_{H}^{2}c_{C}^{2}\,F(m_{H_{1}^{0}}^{2},m_{H_{1}^{\pm}}^{2})+c_{H}^{2}s_{C}^{2}\,F(m_{H_{1}^{0}}^{2},m_{H_{2}^{\pm}}^{2})+s_{H}^{2}c_{C}^{2}\,F(m_{H_{2}^{0}}^{2},m_{H_{1}^{\pm}}^{2})\right.\nonumber \\
 & +s_{H}^{2}s_{C}^{2}\,F(m_{H_{2}^{0}}^{2},m_{H_{2}^{\pm}}^{2})+c_{A}^{2}c_{C}^{2}F(m_{A_{1}^{0}}^{2},m_{H_{1}^{\pm}}^{2})+c_{A}^{2}s_{C}^{2}F(m_{A_{1}^{0}}^{2},m_{H_{2}^{\pm}}^{2})+s_{A}^{2}c_{C}^{2}F(m_{A_{1}^{0}}^{2},m_{H_{1}^{\pm}}^{2})\nonumber \\
 & +s_{A}^{2}s_{C}^{2}F(m_{A_{1}^{0}}^{2},m_{H_{2}^{\pm}}^{2})-c_{H}^{2}c_{A}^{2}F(m_{H_{1}^{0}}^{2},m_{A_{1}^{0}}^{2})-c_{H}^{2}s_{A}^{2}F(m_{H_{1}^{0}}^{2},m_{A_{2}^{0}}^{2})-s_{H}^{2}c_{A}^{2}F(m_{H_{2}^{0}}^{2},m_{A^{0}}^{2})\nonumber \\
 & \left.-s_{H}^{2}s_{A}^{2}F(m_{H_{2}^{0}}^{2},m_{A_{2}^{0}}^{2})\right\} ,\nonumber \\
\varDelta S & =\frac{1}{24\pi}\left\{ (2s_{\mathrm{w}}^{2}-1)^{2}\left[c_{C}^{4}G(m_{H_{1}^{\pm}}^{2},m_{H_{1}^{\pm}}^{2},m_{Z}^{2})+s_{C}^{4}G(m_{H_{2}^{\pm}}^{2},m_{H_{2}^{\pm}}^{2},m_{Z}^{2})+2c_{C}^{2}s_{C}^{2}G(m_{H_{1}^{\pm}}^{2},m_{H_{2}^{\pm}}^{2},m_{Z}^{2})\right]\right.\nonumber \\
 & +c_{H}^{2}c_{A}^{2}\,G(m_{H_{1}^{0}}^{2},m_{A_{1}^{0}}^{2},m_{Z}^{2})+c_{H}^{2}s_{A}^{2}\,G(m_{H_{1}^{0}}^{2},m_{A_{2}^{0}}^{2},m_{Z}^{2})+s_{H}^{2}c_{A}^{2}\,G(m_{H_{2}^{0}}^{2},m_{A_{1}^{0}}^{2},m_{Z}^{2})\nonumber \\
 & +s_{H}^{2}s_{A}^{2}\,G(m_{H_{2}^{0}}^{2},m_{A_{2}^{0}}^{2},m_{Z}^{2})+c_{H}^{2}c_{C}^{2}\,\log\Big(\tfrac{m_{H_{1}^{0}}^{2}}{m_{H_{1}^{\pm}}^{2}}\Big)+c_{H}^{2}s_{C}^{2}\,\log\Big(\tfrac{m_{H_{1}^{0}}^{2}}{m_{H_{2}^{\pm}}^{2}}\Big)+s_{H}^{2}c_{C}^{2}\,\log\Big(\tfrac{m_{H_{2}^{0}}^{2}}{m_{H_{1}^{\pm}}^{2}}\Big)\nonumber \\
 & +s_{H}^{2}s_{C}^{2}\,\log\Big(\tfrac{m_{H_{2}^{0}}^{2}}{m_{H_{2}^{\pm}}^{2}}\Big)+c_{H}^{2}c_{A}^{2}\log\Big(\tfrac{m_{A_{1}^{0}}^{2}}{m_{H_{1}^{\pm}}^{2}}\Big)+c_{H}^{2}s_{A}^{2}\log\Big(\tfrac{m_{A_{2}^{0}}^{2}}{m_{H_{1}^{\pm}}^{2}}\Big)+s_{H}^{2}c_{A}^{2}\log\Big(\tfrac{m_{A_{1}^{0}}^{2}}{m_{H_{2}^{\pm}}^{2}}\Big)\nonumber \\
 & \left.+s_{H}^{2}s_{A}^{2}\log\Big(\tfrac{m_{A_{2}^{0}}^{2}}{m_{H_{2}^{\pm}}^{2}}\Big)\right\} ,\label{dSdT}
\end{align}

where $s_{\mathrm{W}}\equiv\sin~\theta_{W}$, with $\theta_{W}$ is
the Weinberg mixing angle, and $F(x,y)$ and $G(x,y,z)$ are one-loop
functions that can be found in~\cite{Grimus:2008nb}.
\item \textbf{The di-photon Higgs decay}: the Higgs couplings to the two
charged scalar can change drastically the value of the Higgs boson
loop-induced decay into two photons. These loop interactions depends
on the signs and the values of $\omega_{1,2}$, as well on the mixing
and $\theta_{C}$. The ratio $R_{\gamma\gamma}=\mathcal{B}(h\to\gamma\gamma)/\mathcal{B}(h\to\gamma\gamma)^{\mathrm{SM}}=1.02_{-0.12}^{+0.09}$~\cite{ATLAS:2017ovn},
that characterizes the di-photon Higgs decay modification can be written
as
\begin{align}
R_{\gamma\gamma} & =\big(1-\mathcal{B}_{BSM}\big)\left|1+\frac{\upsilon^{2}}{2}\frac{\sum_{j=1}^{2}\frac{r_{j}}{m_{H_{j}^{+}}^{2}}A_{0}^{\gamma\gamma}\big(\frac{m_{h}^{2}}{4m_{H_{j}^{+}}^{2}}\big)}{A_{1}^{\gamma\gamma}\big(\frac{m_{h}^{2}}{4m_{W}^{2}}\big)+\frac{4}{3}A_{1/2}^{\gamma\gamma}\big(\frac{m_{h}^{2}}{4m_{t}^{2}}\big)}\right|^{2},\label{eq:Rgg}
\end{align}
with $r_{1}=c_{C}^{2}\omega_{1}+s_{C}^{2}\omega_{2},\,r_{2}=s_{C}^{2}\omega_{1}+c_{C}^{2}\omega_{2}$;
and $\mathcal{B}_{BSM}$ represents any non-SM decay for the Higgs
like $h\rightarrow H_{i}^{0}H_{k}^{0},\,A_{i}^{0}A_{k}^{0}$. The
loop functions $A_{0,1/2,1}^{\gamma\gamma}$ are given in the literature~\cite{Djouadi:2005gi}.
The $R_{\gamma\gamma}$ ratio estimation in (\ref{eq:Rgg}) is based
on the assumption of the SM production rates for the SM Higgs boson~\cite{ATLAS:2018doi}.
Similarly, the ratio $R_{\gamma Z}=\mathcal{B}(h\to\gamma Z)/\mathcal{B}(h\to\gamma Z)^{\mathrm{SM}}$
can be written as
\begin{align}
R_{\gamma Z} & =\big(1-\mathcal{B}_{BSM}\big)\left|1-\frac{\upsilon^{2}}{2}\frac{1-2s_{W}^{2}}{c_{W}}\frac{\sum_{j=1}^{2}\frac{r_{j}}{m_{H_{j}^{+}}^{2}}A_{0}^{\gamma Z}\big(\frac{m_{h}^{2}}{4m_{H_{j}^{+}}^{2}},\frac{m_{Z}^{2}}{4m_{H_{j}^{+}}^{2}}\big)}{c_{W}~A_{1}^{\gamma Z}\big(\frac{m_{h}^{2}}{4m_{W}^{2}},\frac{m_{Z}^{2}}{4m_{W}^{2}}\big)+\frac{6-16s_{W}^{2}}{3c_{W}}A_{1/2}^{\gamma Z}\big(\frac{m_{h}^{2}}{4m_{t}^{2}},\frac{m_{Z}^{2}}{4m_{t}^{2}}\big)}\right|^{2},\label{eq:Rgz}
\end{align}
with $c_{W}=\cos\theta_{W},~s_{W}=\sin\theta_{W}$ and $\theta_{W}$
is the Weinberg mixing angle; and the loop functions $A_{0,1/2,1}^{\gamma Z}$
are given in the literature~\cite{Djouadi:2005gi}.
\end{itemize}
In Appendix \ref{sec:nInert}, we discuss the generalization of these
results and constraints in the case of a $n$ inert scalar doublets
model.

\section{Viable Parameter Space\label{sec:NumAn}}

In our analysis, we consider both cases where the DM candidate could
be a Majorana fermion $N_{1}$ and the light inert scalar $H_{1}^{0}$.
By taking into account all the above mentioned theoretical and experimental
constraints, we perform a full numerical scan over the model free
parameters (\ref{eq:parfree}). In our scan, we consider the following
ranges for the free parameters
\begin{align}
|s_{H,A,C}|\leq1,~|\omega_{1,2}|<4\pi,~m_{H_{1}^{0},A_{1}^{0}}<3~\textrm{TeV},~78~\textrm{GeV}<m_{H_{1}^{\pm}}<3~\textrm{TeV},
\end{align}
while focusing on the parameter space regions that makes this model
different than a two inert doublets model SM extension, i.e., the
case with non negligible new Yukawa couplings $\min|h_{\alpha i}|\geq10^{-3}$.

\begin{figure}[h]
\includegraphics[width=0.49\textwidth]{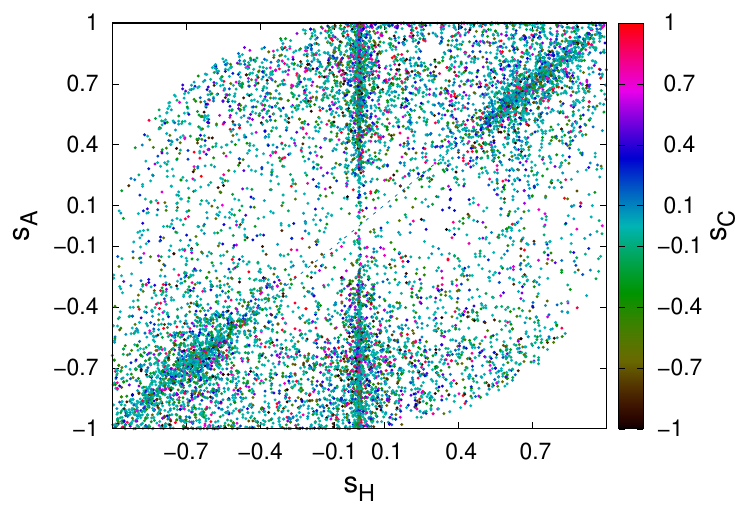}~\includegraphics[width=0.49\textwidth]{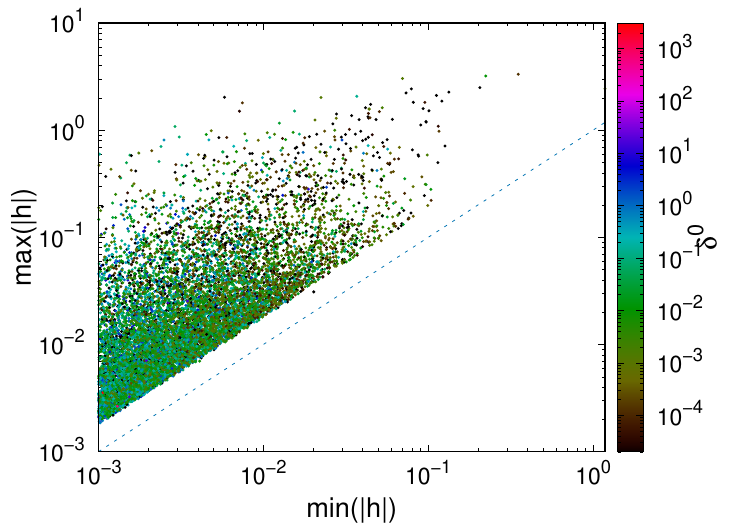}\\
 \includegraphics[width=0.49\textwidth]{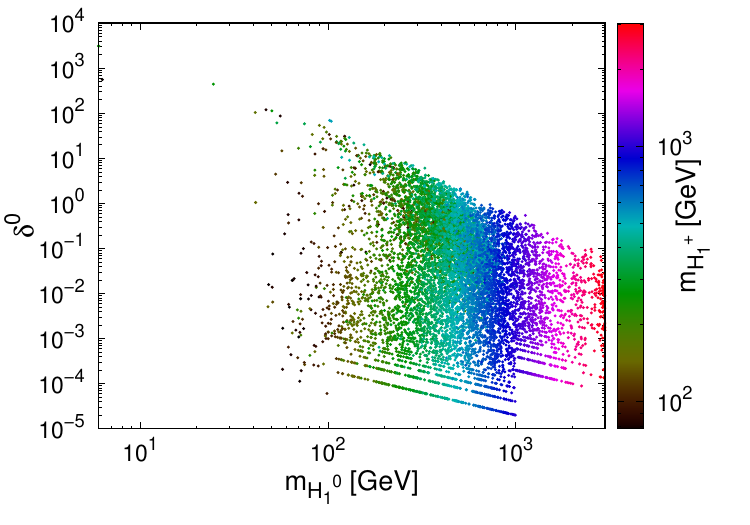}~\includegraphics[width=0.49\textwidth]{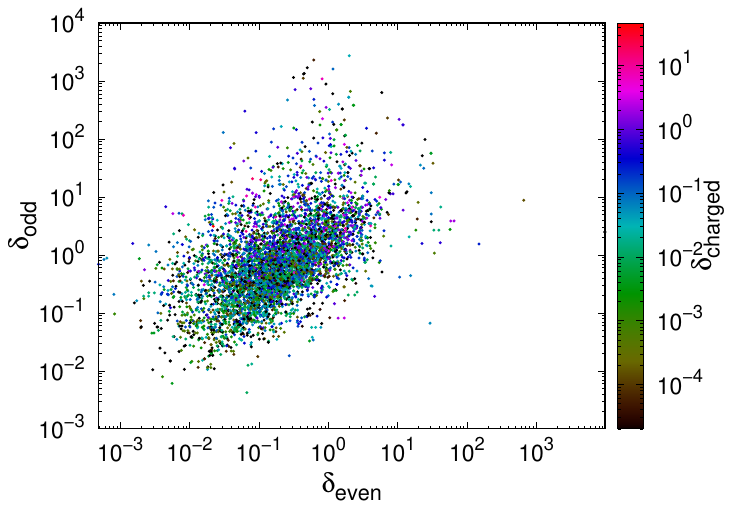}
\caption{Left-top: the CP-even mixing angle sine vs the corresponding CP-odd
one, where the palette shows the mixing of the charged scalar sector.
Right-top: the strength of the new Yukawa couplings $h_{\alpha i}$
where the palette shows the ratio $\delta^{0}=(m_{A_{1}^{0}}^{2}-m_{H_{1}^{0}}^{2})/m_{H_{1}^{0}}^{2}$.
Left-bottom: the mass ratio $\delta^{0}=(m_{A_{1}^{0}}^{2}-m_{H_{1}^{0}}^{2})/m_{H_{1}^{0}}^{2}$
versus the light CP-even scalar mass $m_{H_{1}^{0}}^{2}$, where the
palette shows the light charged scalar mass $m_{H_{1}^{\pm}}^{2}$.
Right-bottom: the ratios $\delta_{even}=(m_{H_{2}^{0}}^{2}-m_{H_{1}^{0}}^{2})/m_{H_{1}^{0}}^{2}$,
$\delta_{even}=(m_{A_{2}^{0}}^{2}-m_{A_{1}^{0}}^{2})/m_{A_{1}^{0}}^{2}$
and $\delta_{charged}=(m_{H_{2}^{\pm}}^{2}-m_{H_{1}^{\pm}}^{2})/m_{H_{1}^{\pm}}^{2}$
that represent the mass relative difference in each sector.}
\label{fig:pr}
\end{figure}

In Fig.~\ref{fig:pr}, using 4k BPs for each (scalar and Majorana)
DM case, we show the mixing angles sine (left-up), the strength of
the new Yukawa couplings $h_{\alpha i}$ (right-up) the masses; and
the inert masses (bottom).

Clearly from Fig.~\ref{fig:pr}, the mixing angles of the CP-even,
CP-odd and charged sectors can take most of the possible ranges, however,
there some preferred values around $s_{H}=0$ and $s_{A}=s_{H}$.
All the above mentioned constraints together with the neutrino oscillation
data that are taken into account via (\ref{eq:h}); lead to the new
Yukawa couplings $h_{\alpha i}$ strength as shown in Fig.~\ref{fig:pr}-top-right.
It is clear that all the couplings can suppressed which makes the
LFV constraints easily fulfilled and the model indistinguishable from
the SM extended by two inert singlets. As mentioned previously, here
we are interested in the opposite case where $|h_{\alpha i}|\geq10^{-3}$,
then, the couplings can be large as $|h_{\alpha i}|<\sqrt{4\pi}$
and a significant hierarchy between the largest and smallest couplings
strength does exist as $\max(|h_{\alpha i}|)>1.8~\min(|h_{\alpha i}|)$.
According to Fig.~\ref{fig:pr}-bottom, the masses of the inert eigenstates
lie within large values intervals.

\begin{figure}[h]
\includegraphics[width=0.49\textwidth]{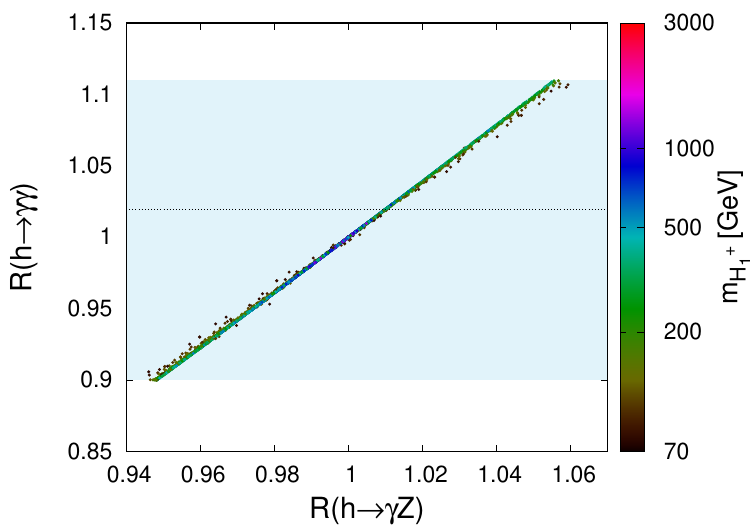}~\includegraphics[width=0.49\textwidth]{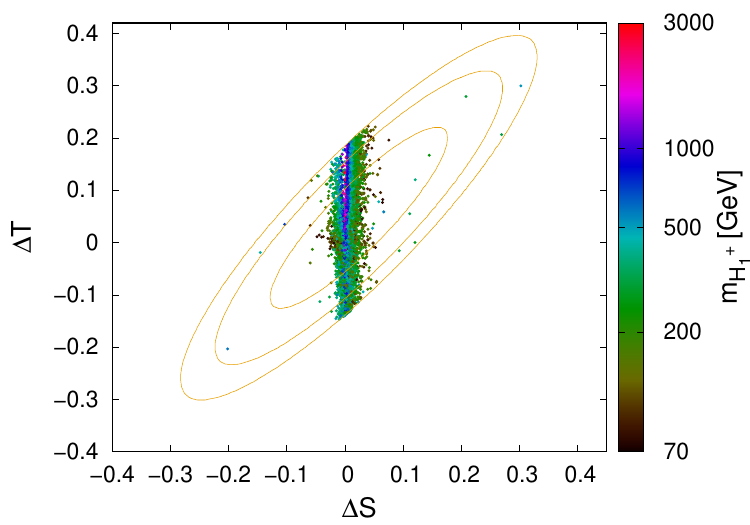}
\caption{Left: The Higgs decay modifiers of $h\rightarrow\gamma\gamma,\gamma\,Z$,
where the colored band represents the allowed experimental values.
Right: the oblique parameters $\Delta~T$ vs $\Delta~S$, where the
palette shows the light charged scalar mass.}
\label{fig:Const}
\end{figure}

In Fig.~\ref{fig:Const}, we present the Higgs decay modifiers of
$h\rightarrow\gamma\gamma,\gamma\,Z$ (left) and the oblique parameters
(right).

One notices that for allowed values for the Higgs branching ratio
$h\rightarrow\gamma\gamma$, the branching ratio $h\rightarrow\gamma\,Z$
can be modified with respect to the SM within the range $[-5.2\%,\,5,1\%]$.
Concerning the electroweak precision tests, as it is expected the
$T$ parameter is more sensitive to the quantum corrections due to
the interactions of the gauge bosons to the two extra doublets, while
the $S$ parameter is sensitive to the corrections only in case of
light charged scalars.

\begin{figure}[h]
\includegraphics[width=0.33\textwidth]{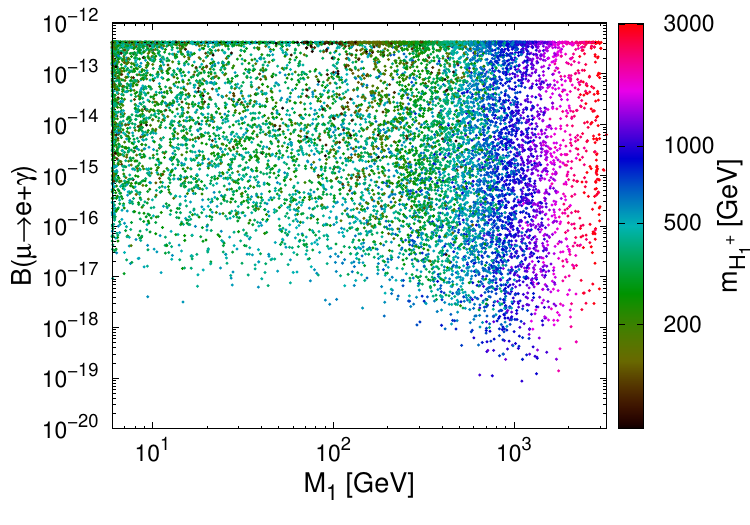}~\includegraphics[width=0.33\textwidth]{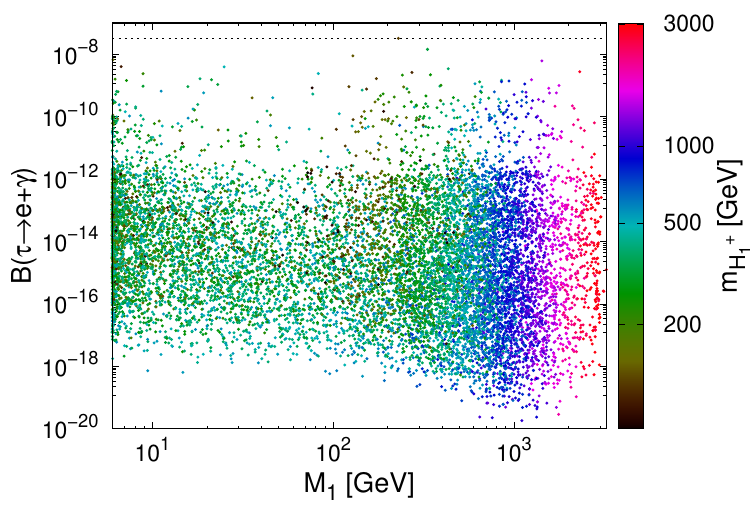}~\includegraphics[width=0.33\textwidth]{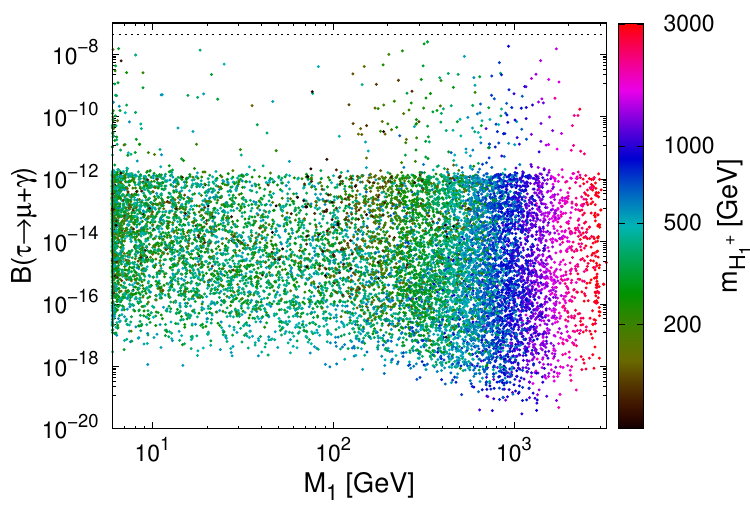}
\caption{The branching ratio of the LFV processes $\mu\rightarrow e+\gamma$,
$\tau\rightarrow e+\gamma$ and $\tau\rightarrow\mu+\gamma$ versus
the light Majorana singlet fermion, where the palette shows the light
charged scalar mass. The dashed horizontal line in each panel represents
the branching ratio experimental bounds.}
\label{fig:LFV}
\end{figure}

In Fig.~\ref{fig:LFV}, we show the branching ratios for the LFV
processes $\ell_{\alpha}\rightarrow\ell_{\beta}+\gamma$ compared
by their experimental bounds.

Like in many models, the constraint form the branching ratio $\mu\rightarrow e\gamma$
is the most severe one and the bounds from the other LFV decays ($\tau\rightarrow e\gamma$
and $\tau\rightarrow\mu\gamma$) are at least three orders of magnitude
smaller.

At colliders, the model may predict many interesting signatures. For
instance, it predicts all the signatures relevant to the MSctM~\cite{Ahriche:2017iar}.
At the LHC, charged scalars can be pair produced as $pp\rightarrow H_{1,2}^{+}H_{1,2}^{-}$
or associated with neutral inert scalars $pp\rightarrow H_{1,2}^{\pm}H_{1,2}^{0},H_{1,2}^{\pm}A_{1,2}^{0}$.
These channels can be probed via the final states mono-lepton $\ell+\slashed{E}_{T}$
and di-lepton $2\ell+\slashed{E}_{T}$ in case if the channels $H_{1}^{\pm}\rightarrow W^{\pm}H_{1,2}^{0},W^{\pm}A_{1,2}^{0}$
are closed. If these channels are open, then the final states $4~jets+\slashed{E}_{T}$
and $1\ell+2~jets+\slashed{E}_{T}$ could be useful to probe this
model as it is useful to probe the MSctM~\cite{Ahriche:2017iar}.
In order to look for beyond MSctM signatures, one has to consider
the heavy inert scalar production at the LHC: $pp\rightarrow H_{2}^{\pm}H_{1,2}^{\mp},H_{2}^{\pm}H_{1,2}^{0},H_{2}^{\pm}A_{1,2}^{0},$
$H_{2}^{0}H_{1,2}^{\pm},A_{2}^{0}H_{1,2}^{\pm},H_{2}^{0}H_{1,2}^{0},A_{2}^{0}H_{1,2}^{0},A_{2}^{0}A_{1,2}^{0}$,
where the heavy scalars should decay as $H_{2}^{+}\rightarrow W^{+}H_{1}^{0},~W^{+}A_{1}^{0},~hH_{1}^{+}$,
$H_{2}^{0}\rightarrow ZA_{1}^{0},~hH_{1}^{0}$ and $A_{2}^{0}\rightarrow ZH_{1}^{0},~hA_{1}^{0}$.
These novel signatures can be probed through the final states $8~jets+\slashed{E}_{T}$,
$1\ell+4~jets+\slashed{E}_{T}$, $1\ell+2~jets+\slashed{E}_{T}$,
$4b+\slashed{E}_{T}$, $2~jets+2b+\slashed{E}_{T}$, $2b+\slashed{E}_{T}$,
that may not be seen in the MSctM or in other neutrino mass and DM
motivated SM extensions. The investigation of these novel signatures
requires a full and precise numerical scan to define the relevant
regions of the parameter space, that should be confronted with the
existing analyses.

\section{Dark Matter: Majorana or Scalar?\label{sec:DM}}

In this model, DM candidate could be either the light Majorana fermion
($N_{1}$), the light $CP$-even ($CP$-odd) scalar $H_{1}^{0}$ ($A_{1}^{0}$),
or a mixture of all these components if they are degenerate in mass.
In the case of scalar DM, the possible annihilation channels are $W^{\pm}W^{\mp}$,~$ZZ$,~$q_{i}\bar{q}_{i}$,
$hh$, $\bar{\ell}_{\alpha}\ell_{\beta}$ and $\bar{\nu}_{\alpha}\nu_{\beta}$.
In this case, very small couplings $h_{\alpha i}$ could be favored
by the LFV constraints, and therefore the contributions of the channels
$\nu_{\alpha}\bar{\nu}_{\beta},\ell_{\alpha}\ell_{\beta}$ to the
annihilation across section would be negligible. Then, this case is
almost identical to the SM extended by two inert doublets with scalar
DM. The co-annihilation effect along the channels $H_{1}^{0}A_{1}^{0}(A_{1}^{0}A_{1}^{0})\rightarrow X_{SM}X'_{SM}$
could be important if the mass difference $(m_{A_{1}^{0}}^{2}-m_{H_{1}^{0}}^{2})/(m_{A_{1}^{0}}^{2}+m_{H_{1}^{0}}^{2})$
is small enough. This makes the couplings $h_{\alpha i}$ non-suppressed
and therefore the annihilation into $\nu_{\alpha}\bar{\nu}_{\beta},\ell_{\alpha}\ell_{\beta}$
is more important. Such effect could make this setup (spin-$0$ DM)
better than the minimal IDM. In case of Majorana DM scenario, the
DM (co-)annihilation could occur into charged leptons $\ell_{\alpha}^{-}\ell_{\beta}^{+}$
and light neutrinos $\nu_{\alpha}\bar{\nu}_{\beta}$; via $t$-channel
diagrams mediated by the charged scalar $H_{1,2}^{\pm}$; and the
neutral scalars $H_{1,2}^{0},A_{1,2}^{0}$, respectively. Here, the
annihilation cross section is fully triggered by the non-suppressed
values of the couplings $h_{\alpha i}$.

\begin{figure}[h]
\begin{centering}
\includegraphics[width=0.55\textwidth]{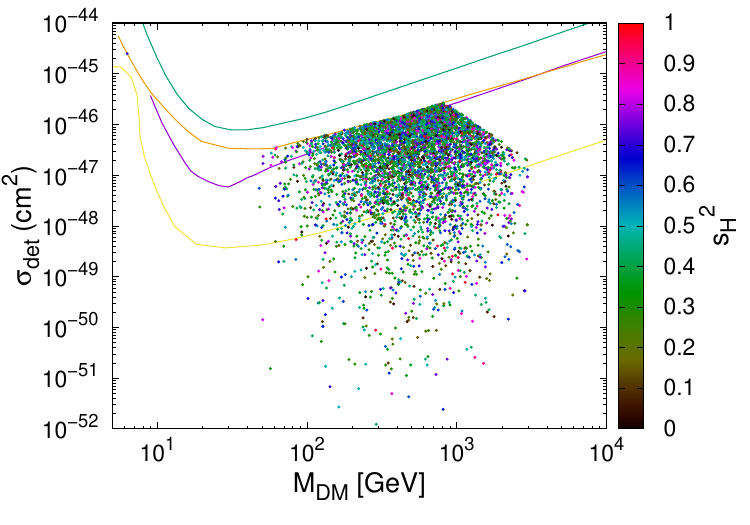}
\par\end{centering}
\caption{The DD cross section vs the DM mass for the scalar DM BPs used in
Fig.~\ref{fig:pr}. The violet, blue and orange lines correspond
to the bounds from LUX-ZEPLIN~\cite{LUX-ZEPLIN:2022qhg}, Xenon-1T
bound~\cite{XENON:2015gkh} and PandaX-4T~\cite{PandaX-4T:2021bab},
respectively, while the yellow one represents the neutrino floor~\cite{Billard:2013qya}.}
\label{fig:DM}
\end{figure}

In case of scalar DM, negative searches from DM direct detection (DD)
at underground detectors impose more constraints on the parameters
space. This is translated into upper bounds on the DM-nucleon cross
section. In our model, the DM interaction with nucleons occurs via
a single t-channel diagram mediated by the Higgs, and the scattering
cross-section is given by\footnote{In the scalar DM scenario, the vertex $ZH_{1}^{0}H_{1}^{0}$ is vanishing
since violates CP symmetry. However, the vertex $ZH_{1}^{0}A_{1}^{0}$
does not vanish, but the process $H_{1}^{0}\mathcal{N}\rightarrow A_{1}^{0}\mathcal{N}$
is kinematically forbidden. In case of Majorana DM, the vertex $hN_{1}N_{1}$
($ZN_{1}N_{1}$) is vanishing at tree-level due to the fact that the
fermion $N_{1}$ is a singlet (a Majorana one). At loop level, it
has been shown that DD cross section is few orders of magnitude suppressed~\cite{Ahriche:2016acx,Ahriche:2017iar}.}
\begin{equation}
\sigma_{det}=\frac{(\lambda_{hH_{1}^{0}H_{1}^{0}})^{2}}{4\pi\upsilon^{2}m_{h}^{4}}\frac{m_{\mathcal{N}}^{2}(m_{\mathcal{N}}-\frac{7}{9}m_{\mathcal{B}})^{2}}{(m_{\mathcal{N}}+m_{H_{1}^{0}})^{2}},\label{eq:DD}
\end{equation}
where $m_{\mathcal{N}}$ and $m_{\mathcal{B}}$ are, respectively,
the nucleon and baryon masses in the chiral limit \cite{He:2008qm};
and $\lambda_{hH_{1}^{0}H_{1}^{0}}=\frac{\upsilon}{2}[c_{H}^{2}(\kappa_{1}+\omega_{1}+\xi_{1})+s_{H}^{2}(\kappa_{2}+\omega_{2}+\xi_{2})+2s_{H}c_{H}(\xi_{3}+\xi_{4})]$
is the DM Higgs triple coupling. In Fig.~\ref{fig:DM}, we show the
DD cross section (\ref{eq:DD}) versus the DM mass for the BPs used
in Fig.~\ref{fig:pr} that correspond to scalar DM cases, compared
with recent experimental bounds from the PandaX-4T experiment~\cite{PandaX-4T:2021bab}
and LUX-ZEPLIN~\cite{LUX-ZEPLIN:2022qhg}.

In order to show that both scalar and Majorana DM scenarios are viable
in this model, we consider the two BPs among the BPs used in Fig.~\ref{fig:pr},
where their free parameters values are given Table~\ref{BPs}, where
BP1 and BP2 correspond to Majorana and scalar DM cases, respectively.

\begin{table}[h]
\begin{centering}
\resizebox{\textwidth}{!}{%
\begin{tabular}{|c||c|}
\hline
BP1  & $s_{H,A,C}=\{-0.053734,~0.72187,~0.82912\},~\omega_{1,2}=\{-0.15715,~-0.92515\}$,\tabularnewline
 & $m_{H_{i}^{0}}=\{2755.6,2820.6\},\,m_{A_{i}^{0}}=\{2766.7,2823\}$,\tabularnewline
 & $m_{H_{i}^{\pm}}=\{2318.7,~2327.8\},\,M_{i}=\{2317.1,~2318.7,~2327.8\}$,\tabularnewline
 & $h_{\alpha i}=\left(\begin{array}{cccccc}
-1.3664-i0.066927, & -0.0063156-i1.1746, & -0.020184+i0.28501, & -0.0063503-i1.1811, & -1.3739-i0.067294, & -0.020295+i0.28657\\
1.8625+i0.046833, & -0.09083-i2.2339, & -0.0063982-i0.33045, & -0.091329-i2.2461, & 1.8727+i0.04709, & -0.0064333-i0.33226\\
-2.4404-i0.065694, & -0.667-i1.8413, & 0.037294-i0.36281, & -0.67066-i1.8514, & -2.4538-i0.066055, & 0.037499-i0.36481
\end{array}\right)$ \tabularnewline
\hline
\hline
BP2  & $s_{H,A,C}=\{-0.38546,\,-0.27092,\,-0.019603\},~\omega_{1,2}=\{-1.0362,\,5.1668\}$,\tabularnewline
 & $m_{H_{i}^{0}}=\{1228.9,~1246\},\,m_{A_{i}^{0}}=\{1229.3,~1266\}$,\tabularnewline
 & $m_{H_{i}^{\pm}}=\{1230.4,~1270.1\},\,M_{i}=\{1633.4,~1850.7,~1853.5\}$,\tabularnewline
 & $h_{\alpha i}=10^{-3}\times\left(\begin{array}{cccccc}
2.5475-i51.865, & 15.027+i1.4322, & 19.457+i0.69002, & 0.44292-i9.0175, & 2.6139+i0.24914, & 3.3845+i0.12003\\
-1.7657+i70.885, & -31.844+i0.82879, & 4.9703-i0.46673, & -0.307+i12.325, & -5.5394+i0.14417, & 0.86461-i0.08119\\
2.5076-i92.163, & -33.608+i0.041734, & 1.0967-i6.4609, & 0.43599-i16.024, & -5.8463+i0.0072598, & 0.19077-i1.1239
\end{array}\right)$ \tabularnewline
\hline
\end{tabular}}
\par\end{centering}
\caption{The values of the free parameters for BP1 and BP2. Here, all the masses
are given in GeV.}
\label{BPs}
\end{table}

\begin{figure}[h]
\begin{centering}
\includegraphics[width=0.49\textwidth]{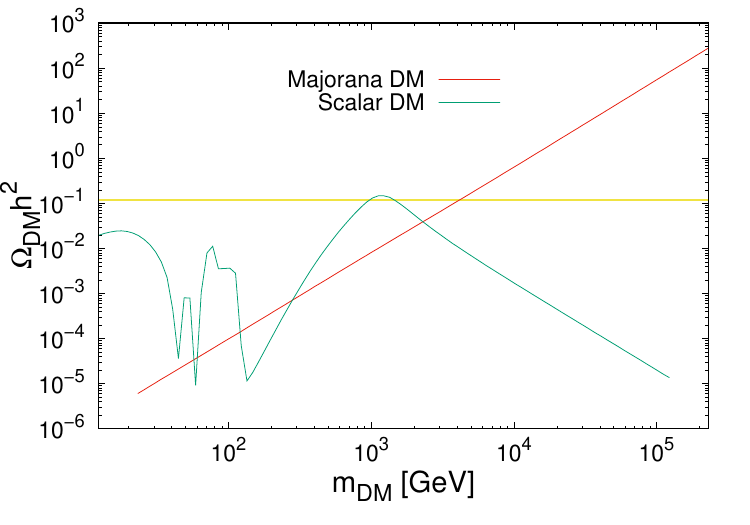}~\includegraphics[width=0.49\textwidth]{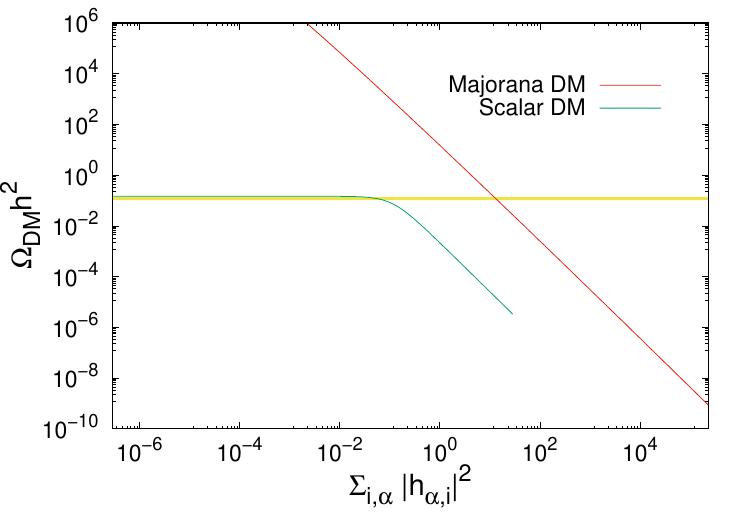}
\par\end{centering}
\caption{The DM relic density versus the DM mass (left) and versus the couplings
combination $\sum_{i,\alpha}|h_{\alpha,i}|^{2}$ (right). The yellow
band represents the observed the DM relic density value. One has to
mention that in the left panel, we make all masses changing simultaneously,
i.e., all the mass fulfill the condition $m_{i}/m_{DM}=constant$
with all the DM mass interval. }
\label{fig:Omh}
\end{figure}

Then, we use the package MadDM~\cite{Ambrogi:2018jqj} to estimate
the relic density and the direct detection cross section for the scalar
DM case by varying both the DM mass and the Yukawa couplings $h_{\alpha i}$
strength as shown in Fig.~\ref{fig:Omh}. In order to do so, we used
the package FeynRules~\cite{Alloul:2013bka} to implement the model
and produce the UFO files~\cite{UFO}.

According to Fig.~\ref{fig:Omh}, it is clear that the right amount
of the DM relic density can be achieved in this model whether the
DM is a Majorana or a scalar particle. From Fig.~\ref{fig:Omh}-left,
the dependence of the relic density on the DM mass in case of Majorana
DM seems to be quadratic $\Omega_{DM}h^{2}\sim m_{DM}^{2}$ (or $\sigma_{DM}\sim1/m_{DM}^{2}$).
This can be understood due to the fact that the Majorana DM (co-)annihilation
occurs via simple channels, i.e., $N_{1}N_{1}(N_{2,3})\rightarrow\nu_{\alpha}\bar{\nu}_{\beta},\ell_{\alpha}\ell_{\beta}$.
However, in the scalar DM case, its dependence on the DM mass is not
trivial due to the existence of many annihilation channels ($N_{1}N_{1}(N_{2,3})\rightarrow W^{\pm}W^{\mp}$,~$ZZ$,~$q_{i}\bar{q}_{i}$,
$hh$, $\bar{\ell}_{\alpha}\ell_{\beta}$ and $\bar{\nu}_{\alpha}\nu_{\beta}$),
where each channel could be dominant within a specific DM mass range.
From Fig.~\ref{fig:Omh}-right, one learns that in the scalar DM
case the DM annihilation into $\nu_{\alpha}\bar{\nu}_{\beta},\ell_{\alpha}\ell_{\beta}$
could be efficient only if $\sum_{i,\alpha}|h_{\alpha,i}|^{2}\sim0.1$.
However, since the DM is annihilated only via the channels $\nu_{\alpha}\bar{\nu}_{\beta},\ell_{\alpha}\ell_{\beta}$
for the Majorana DM case, the Yukawa coupling magnitude is very important,
and it is solely dictated by the relic density right amount. Therefore,
Yukawa couplings of order ${\cal O}(0.1\sim1)$ could be problematic
for the LFV constraints. In order to fulfill the requirements of DM
relic density, the LFV constraints and the neutrino oscillation data
together, one has to impose that the six terms in the summation in
(\ref{LFV}) should cancel together. Although, this is not surprising
since it has been shown in the MSctM with Majorana DM~\cite{Ahriche:2017iar}
that these requirements are fulfilled by the imposing the cancellation
of three terms in a formula similar to (\ref{LFV}); for a large DM
mass range. Indeed, in our model, the Feynman diagrams relevant to
the neutrino mass, LFV and DM annihilation are more numerous; and
involve more free parameters with respect the MSctM, and hence make
the DM relic density, the LFV constraints and the neutrino oscillation
data easily relaxed with respect to the case in~\cite{Ahriche:2017iar}.

\section{Conclusion~\label{sec:conclusion}}

In this paper, we proposed a SM extension by three singlet Majorana
fermions and two scalar inert doublets. After diagonalizing the scalar
squared mass sub-matrices, we found that the model includes two CP-even
($H_{1,2}^{0}$), two CP-odd ($A_{1,2}^{0}$) and two charged scalar
($H_{1,2}^{\pm}$) eigenstates, with different mixing angles. Then,
the values of many physical observables got modified with respect
to the MSctM. For instance, the neutrino mass is generated at one-loop
level via 12 diagrams instead of 6 in the MSctM. Similar remarks hold
for the observables relevant to the electroweak precision tests, LFV
and the di-photon Higgs decay, where many masses and mixing angles
are included. This made the neutrino oscillation data easily explained
without being in conflict with the different theoretical and experimental
constraints for large parameter space regions.

The dark matter (DM) in this model could be either fermionic ($N_{1}$)
or scalar (the lightest among $H_{1}^{0}$ and $A_{1}^{0}$). We have
shown that the right amount of the relic density can be easily achieved
in both scenarios for different DM mass ranges. In addition, for the
scalar DM scenario the model can accommodate the constraints for the
direct detection cross section of the DM-nucleon scattering.

The model predicts all the signatures predicted by the MSctM, however,
novel signature relevant to the production and decay of the heavy
scalars ($H_{2}^{\pm},H_{2}^{0},A_{2}^{0}$) can be probed through
some final states like: $8~jets+\slashed{E}_{T}$, $1\ell+4~jets+\slashed{E}_{T}$,
$1\ell+2~jets+\slashed{E}_{T}$, $4b+\slashed{E}_{T}$, $2~jets+2b+\slashed{E}_{T}$
and $2b+\slashed{E}_{T}$. This point requires an independent investigation.

\acknowledgments

This work is funded by the University of Sharjah under the research
projects No 21021430100 ``\textit{Extended Higgs Sectors at Colliders:
Constraints \& Predictions}'' and No 21021430107 ``\textit{Hunting
for New Physics at Colliders}''.

\appendix

\section{Generalized Casas-Ibarra parameterization~\label{app1}}

It is well known that a (complex) $3\times3$ fermionic mass matrices
can be diagonalized via two unitary matrices $U$ and $V$ à la singular
decomposition value problem~\cite{SingDec}, as
\begin{equation}
m^{(\nu)}=U.D_{m}.V^{\dagger},\label{eq:nu}
\end{equation}
with $D_{m}=\mathrm{diag\big\{ m_{1},m_{2},m_{3}\big\}}$. Here, $U_{3\times3}$
and $V_{3\times3}$ can be defined as the unitary matrices that diagonalize
the matrices products $m^{(\nu)}.(m^{(\nu)})^{\dagger}$ and $(m^{(\nu)})^{\dagger}.m^{(\nu)}$,
respectively. In other words $U^{\dagger}.[m^{(\nu)}.(m^{(\nu)})^{\dagger}].U=D_{m}^{\dagger}.D_{m}=D_{m^{2}}=D_{m}.D_{m}^{\dagger}=V^{\dagger}.[(m^{(\nu)})^{\dagger}.m^{(\nu)}].V$.
In the case of neutrino Majorana mass, the matrix (\ref{eq:nu}) is
complex symmetric, which implies $U=V^{*}$ and therefore
\begin{equation}
m^{(\nu)}=U.D_{m}.U^{T}\,\,.\label{eq:MD}
\end{equation}
This allows to write a general form for the diagonal matrix $D_{m}$
as
\begin{align}
D_{m}=\mathcal{P}_{1}.U^{T}.m^{(\nu)}.U.\mathcal{P}_{2}\,\,\, & ,\label{eq:Mnu}
\end{align}
where the matrices $\mathcal{P}_{1,2}=\mathrm{diag}\big\{1,e^{i\zeta_{1,2}},e^{i\vartheta_{1,2}}\big\}$
can be absorbed in a redefinition of the Majorana phase matrix in
the PNMS mixing matrix; and similarly, $D_{m}$ in (\ref{eq:Mnu})
can be converted to a diagonal matrix with real positive elements.
This parameterization (\ref{eq:Mnu}) is in agreement with the so-called
Takagi diagonalization~\cite{Takagi}.

In what follows, we consider the new Yukawa couplings are given by
the matrix $h_{3,M}$, with $M=3n$ and is the inert doublets number
that are considered in the scotogenic model generalization. In this
case, the matrix $\Lambda$ in (\ref{eq:CIb}) should be a $M\times M$
matrix rather than $6\times6$ one. In our case, we have $n=2$ and
then the Yukawa couplings are presented by $3\times6$ complex matrix.
Following the Casas-Ibarra approach~\cite{Casas:2001sr}, the Yukawa
couplings are assumed to have the form of a product of many matrices
as
\begin{equation}
h=S_{1}.S_{2}.S_{3}.S_{4}.S_{5}.\label{eq:hS}
\end{equation}

Then by replacing (\ref{eq:hS}) in (\ref{eq:CIb}); and match it
with (\ref{eq:MD}), one immediately finds that $S_{1}=U_{3\times3}$
is the $3\times3$ PNMS mixing matrix. In order to identify the remaining
matrices, we simplify the product $U^{\dagger}.m^{(\nu)}.(m^{(\nu)})^{\dagger}.U$
and match it with $D_{m^{2}}$. Then, in case where the matrix $S_{5}=Q$
is chosen to be the orthogonal $M\times M$ matrix that diagonalizes
the symmetric real matrix $\Lambda$ as $Q_{M\times M}.(\Lambda)_{M\times M}.Q_{M\times M}^{T}=D_{\Lambda'}=\mathrm{diag}\big\{\Lambda_{1}',\Lambda_{2}',...,\Lambda_{M}'\big\}$;
and the matrix $S_{4}$ to be the diagonal $M\times M$ matrix $S_{4}=D_{(\Lambda')^{-1/2}}=\mathrm{diag}\big\{(\Lambda_{1}')^{-1/2},(\Lambda_{2}')^{-1/2},...,(\Lambda_{M}')^{-1/2}\big\}$,
then the product $U^{\dagger}.m^{(\nu)}.(m^{(\nu)})^{\dagger}.U$
can be written as
\begin{align}
U^{\dagger}.m^{(\nu)}.(m^{(\nu)})^{\dagger}.U & =(S_{2}.S_{3})_{3\times M}.(S_{3}^{T}.S_{2}^{T})_{M\times3}.(S_{2}^{*}.S_{3}^{*})_{3\times M}.(S_{3}^{\dagger}.S_{2}^{\dagger})_{M\times3}.\label{eq:UMMU}
\end{align}
Here, $\Lambda'_{i}$ are the eigenvalues of the matrix $\Lambda$.
In order to match (\ref{eq:UMMU}) with$D_{m^{2}}$, it is enough
to have $S_{2}=D_{m^{1/2}}=\mathrm{diag\big\{ m_{1}^{1/2},m_{2}^{1/2},m_{3}^{1/2}\big\}}$
and $S_{3}=T$ with $(T)_{3\times M}.(T^{T})_{M\times3}=\boldsymbol{1}_{3\times3}$.
Obviously, for $M=3$ the matrix $T$ is just a $3\times3$ orthogonal
matrices $T.T^{T}=T^{T}.T=\boldsymbol{1}_{3\times3}$. However, for
$M=3\times k$, we claim that the matrix $T$ has the most general
form
\begin{eqnarray}
T_{3\times(3k)} & = & \big(\begin{array}{ccccc}
\alpha^{(1)}R_{3\times3}^{(1)}, & \alpha^{(2)}R_{3\times3}^{(2)}, & ....... & . & \alpha^{(k)}R_{3\times3}^{(k)}\end{array}\big),\label{eq:T}
\end{eqnarray}
with $R^{(i)}$ are $3\times3$ orthogonal matrices $R^{(i)}.(R^{(i)})^{T}=(R^{(i)})^{T}.R^{(i)}=\boldsymbol{1}_{3\times3}$,
and $\alpha^{(i)}$ are real numbers that fulfill the identity $\sum_{i}(\alpha^{(i)})^{2}=1$.

In order to get large Yukawa couplings as favored by collider searches
or required by Majorana DM relic density, it is enough to choose a
diagonalization matrix $(Q)_{M\times M}$ that leads to the order
$|\Lambda'_{1}|<|\Lambda'_{2}|<..<<|\Lambda'_{M}|$; and $\alpha^{(1)}=1,\,\alpha^{(i>1)}=0$.

\section{Scotogenic Model with $n$ inert Doublets\label{sec:nInert}}

In case of a scotogenic model with N inert doublets ($\Phi_{i=1,n}$)
with the global $Z_{2}$ symmetry $N_{i}\rightarrow-N_{i},~\Phi_{i}\rightarrow-\Phi_{i}$,
the Lagrangian (\ref{LL}) can be generalized as
\begin{eqnarray}
\mathcal{L} & \supset & \sum_{i=1}^{n}\sum_{k=1}^{3}\overline{L}_{\alpha}\big((h_{\alpha,k+3(i-1)})\epsilon\Phi_{i}\big)N_{k}+\frac{1}{2}\bar{N}_{k}^{C}M_{k}N_{k}+h.c.,\label{LLL}
\end{eqnarray}
where the new Yukawa couplings $h_{\alpha,i}$ here are a $3\times(3n)$
matrix. The scalar potential (\ref{eq:V}) can be generalized as
\begin{eqnarray}
V\left(\mathcal{H},\Phi_{i},S,\chi\right) & = & -\mu_{H}^{2}|\mathcal{H}|^{2}+\mu_{i}^{2}|\Phi_{i}|^{2}+\frac{\lambda_{H}}{6}|\mathcal{H}|^{4}+\frac{\lambda_{i}}{6}|\Phi_{i}|^{4}+\omega_{i}|\mathcal{H}|^{2}|\Phi_{i}|^{2}+\kappa_{i}|\mathcal{H}^{\dagger}\Phi_{i}|^{2}\nonumber \\
 &  & +\varrho_{i,j}^{(1)}|\Phi_{i}|^{2}|\Phi_{j}|^{2}+\varrho_{i,j}^{(2)}|\Phi_{i}^{\dagger}\Phi_{j}|^{2}+\left\{ \mu_{3,ij}^{2}\Phi_{i}^{\dagger}\Phi_{j}+\frac{1}{2}\xi_{i}^{(1)}(\mathcal{H}^{\dagger}\Phi_{i})^{2}\right.\nonumber \\
 &  & \left.+\xi_{i,j}^{(2)}(\mathcal{H}^{\dagger}\Phi_{i})(\mathcal{H}^{\dagger}\Phi_{j})+\xi_{i,j}^{(3)}(\Phi_{i}^{\dagger}\mathcal{H})(\mathcal{H}^{\dagger}\Phi_{j})+h.c.\right\} .\label{eq:VV}
\end{eqnarray}

In case of CP-conservation (real values of all potential parameters),
there are $n$ charged eigenstates ($H_{j=1,n}^{\pm}$), $n$ neutral
CP-even eigenstates ($H_{j=1,n}^{0}$) and $n$ neutral CP-odd eigenstates
($A_{j=1,n}^{0}$) with similar couplings to (\ref{coup}). In (\ref{coup}),
the mixing ($c_{H,A,C}$ and $s_{H,A,C}$ should be replaced by the
elements of the mixing matrices ${\cal U}^{(H,A,C)}$ that diagonalize
the CP-even, CP-odd and charged scalars squared $n\times n$ mass
matrices, respectively, where the summation should be performed over
$j=1,n$. Consequently, the neutrino mass is generated via $4\times n$
diagrams mediated by $N_{1,2,3}$ and $H_{j=1,n}^{0}/A_{j=1,n}^{0}$.
Therefore, the formulas of the mass matrix elements (\ref{eq:mnu})
and the values of the new Yukawa couplings in (\ref{eq:h}) hold for
$j=1,n$ and $M=3\times n$. Similar conclusions can be achieved for
the LFV branching ratios (\ref{LFV}) and the di-photon Higgs decay
ratio (\ref{eq:Rgg}). For the oblique parameters (\ref{dSdT}), the
summation should be performed over $j=1,n$; and the mixing ($c_{H,A,C}$
and $s_{H,A,C}$ should be replaced by the elements of the mixing
matrix ${\cal U}^{(H,A,C)}$. By considering $n_{N}$ Majorana singlet
fermions instead of three, the description of this generalization
corresponds what called the \textit{general scotogenic model}~\cite{Escribano:2020iqq}.

\section{The Unitarity Amplitude Matrices~\label{Uni}}

Here, we give the relevant amplitude matrices for the unitarity. We have the
following basis that characterize the initial/final states:

\textbf{CP-even, $Z_2 $ even, \& $Q_{em}=0$}: the basis is \{$hh$,
$\chi^0\chi^0$, $S_1^0S_1^0$, $S_2^0S_2^0$,
$Q_1^0Q_1^0$, $Q_2^0Q_2^0$, $\chi^{+}\chi^{-}$,
$S_1^{+}S_1^{-}$, $S_2^{+}S_2^{-}$, $S_1^{+}S_2^{-}$\};
and the matrix is
\begin{equation}
\resizebox{0.93\textwidth}{!}{$\left[
\begin{array}{cccccccccc}
\lambda_{H}, & \frac{1}{3}\lambda_{H}, & \omega_1 +\kappa_1 +\xi_1 , & \omega_2 +\kappa_2 +\xi_2 , & \omega_1 +\kappa_1 -\xi_1 , & \omega_2 +\kappa_2 -\xi_2 , & \frac{1}{3}\lambda_{H}, & \omega_1 , & \omega_2 , & 0\\
\frac{1}{3}\lambda_{H}, & \lambda_{H}, & \omega_1 +\kappa_1 -\xi_1 , & \omega_2 +\kappa_2 -\xi_2 , & \omega_1 +\kappa_1 +\xi_1 , & \omega_2 +\kappa_2 +\xi_2 , & \frac{1}{3}\lambda_{H}, & \omega_1 , & \omega_2 , & 0\\
\omega_1 +\kappa_1 +\xi_1 , & \omega_1 +\kappa_1 -\xi_1 , & \lambda_1 , & \varrho_1 +\varrho_2 , & \frac{1}{3}\lambda_1 , & \varrho_1 +\varrho_2 , & \omega_1 , & \frac{1}{3}\lambda_1 , & \varrho_1 , & 0\\
\omega_2 +\kappa_2 +\xi_2 , & \omega_2 +\kappa_2 -\xi_2 , & \varrho_1 +\varrho_2 , & \lambda_2 , & \varrho_1 +\varrho_2 , & \frac{1}{3}\lambda_2 , & \omega_2 , & \varrho_1 , & \frac{1}{3}\lambda_2 , & 0\\
\omega_1 +\kappa_1 -\xi_1 , & \omega_1 +\kappa_1 +\xi_1 , & \frac{1}{3}\lambda_1 , & \varrho_1 +\varrho_2 , & \lambda_1 , & \varrho_1 +\varrho_2 , & \omega_1 , & \frac{1}{3}\lambda_1 , & \varrho_1 , & 0\\
\omega_2 +\kappa_2 -\xi_2 , & \omega_2 +\kappa_2 +\xi_2 , & \varrho_1 +\varrho_2 , & \frac{1}{3}\lambda_2 , & \varrho_1 +\varrho_2 , & \lambda_2 , & \omega_2 , & \varrho_1 , & \frac{1}{3}\lambda_2 , & 0\\
\frac{1}{3}\lambda_{H}, & \frac{1}{3}\lambda_{H}, & \omega_1 , & \omega_2 , & \omega_1 , & \omega_2 , & \frac{2}{3}\lambda_{H}, & \kappa_1 +\omega_1 , & \kappa_2 +\omega_2 , & \xi_{4}\\
\omega_1 , & \omega_1 , & \frac{1}{3}\lambda_1 , & \varrho_1 , & \frac{1}{3}\lambda_1 , & \varrho_1 , & \kappa_1 +\omega_1 , & \frac{2}{3}\lambda_1 , & \varrho_1 +\varrho_2 , & 0\\
\omega_2 , & \omega_2 , & \varrho_1 , & \frac{1}{3}\lambda_2 , & \varrho_1 , & \frac{1}{3}\lambda_2 , & \kappa_2 +\omega_2 , & \varrho_1 +\varrho_2 , & \frac{2}{3}\lambda_2 , & 0\\
0, & 0, & 0, & 0, & 0, & 0, & \xi_{4}, & 0, & 0, & \varrho_1 +\varrho_2
\end{array}\right]$}
\end{equation}

\textbf{CP-even, $Z_2 $ even, \& $Q_{em}=0$}: the basis is \{$hS_1^0$,
$hS_2^0$, $\chi^0Q_1^0$, $\chi^0Q_2^0$, $\chi^{+}S_1^{-}$,
$\chi^{+}S_2^{-}$\}; and the matrix is
\begin{equation}
\resizebox{0.93\textwidth}{!}{$\left[
\begin{array}{cccccc}
\omega_1 +\kappa_1 +\xi_1 , & \xi_{3}+\xi_{4}, & \xi_1 , & \xi_{3}, & \frac{1}{2}(\kappa_1 +\xi_1 ), & \frac{1}{2}(\xi_{3}+\xi_{4})\\
\xi_{3}+\xi_{4}, & \omega_2 +\kappa_2 +\xi_2 , & \xi_{3}, & \xi_2 , & \frac{1}{2}(\xi_{3}+\xi_{4}), & \frac{1}{2}(\kappa_2 +\xi_2 )\\
\xi_1 , & \xi_{3}, & \omega_1 +\kappa_1 +\xi_1 , & \xi_{3}+\xi_{4}, & \frac{1}{2}(\kappa_1 +\xi_1 ), & \frac{1}{2}(\xi_{3}+\xi_{4})\\
\xi_{3}, & \xi_2 , & \xi_{3}+\xi_{4}, & \omega_2 +\kappa_2 +\xi_2 , & \frac{1}{2}(\xi_{3}+\xi_{4}), & \frac{1}{2}(\kappa_2 +\xi_2 )\\
\frac{1}{2}(\kappa_1 +\xi_1 ), & \frac{1}{2}(\xi_{3}+\xi_{4}), & \frac{1}{2}(\kappa_1 +\xi_1 ), & \frac{1}{2}(\xi_{3}+\xi_{4}), & \kappa_1 +\omega_1 , & \xi_{4}\\
\frac{1}{2}(\xi_{3}+\xi_{4}), & \frac{1}{2}(\kappa_2 +\xi_2 ), & \frac{1}{2}(\xi_{3}+\xi_{4}), & \frac{1}{2}(\kappa_2 +\xi_2 ), & \xi_{4}, & \kappa_2 +\omega_2
\end{array}\right]$}
\end{equation}

\textbf{CP-even, $Z_2 $ even, \& $Q_{em}=0$}: the basis is \{$h\chi^0$,
$S_1^0Q_1^0$, $S_1^0Q_2^0$, $S_2^0Q_1^0$,
$S_2^0Q_2^0$, $\chi^{+}\chi^{-}$, $S_1^{+}S_1^{-}$,
$S_2^{+}S_2^{-}$, $S_1^{+}S_2^{-}$\}; and the matrix is
\begin{equation}
\left[\begin{array}{ccccccccc}
\frac{1}{3}\lambda_{H}, & \xi_1 , & \xi_{3}, & \xi_{3}, & \xi_2 , & 0, & 0, & 0, & 0\\
\xi_1 , & \frac{1}{3}\lambda_1 , & 0, & 0, & 0, & 0, & 0, & 0, & 0\\
\xi_{3}, & 0, & \varrho_1 +\varrho_2 , & 0, & 0, & 0, & 0, & 0, & -\frac{i}{2}\varrho_2 \\
\xi_{3}, & 0, & 0, & \varrho_1 +\varrho_2 , & 0, & 0, & 0, & 0, & \frac{i}{2}\varrho_2 \\
\xi_2 , & 0, & 0, & 0, & \frac{1}{3}\lambda_2 , & 0, & 0, & 0, & 0\\
0, & 0, & 0, & 0, & 0, & \frac{2}{3}\lambda_{H}, & \kappa_1 +\omega_1 , & \kappa_2 +\omega_2 , & \xi_{4}\\
0, & 0, & 0, & 0, & 0, & \kappa_1 +\omega_1 , & \frac{2}{3}\lambda_1 , & \varrho_1 +\varrho_2 , & 0\\
0, & 0, & 0, & 0, & 0, & \kappa_2 +\omega_2 , & \varrho_1 +\varrho_2 , & \frac{2}{3}\lambda_2 , & 0\\
0, & 0, & \frac{i}{2}\varrho_2 , & -\frac{i}{2}\varrho_2 , & 0, & \xi_{4}, & 0, & 0, & \varrho_1 +\varrho_2
\end{array}\right]
\end{equation}

\textbf{CP-even, $Z_2 $ even, \& $Q_{em}=0$}: the basis is \{$hQ_1^0$,
$hQ_2^0$, $S_1^0\chi^0$, $S_2^0\chi^0$, $\chi^{+}S_1^{-}$,
$\chi^{+}S_2^{-}$\}; and the matrix is
\begin{equation}
\resizebox{0.93\textwidth}{!}{$\left[
\begin{array}{cccccc}
\omega_1 +\kappa_1 -\xi_1 , & -\xi_{3}+\xi_{4}, & \xi_1 , & \xi_{3}, & -\frac{i}{2}(\kappa_1 -\xi_1 ), & \frac{i}{2}(\xi_{3}-\xi_{4})\\
-\xi_{3}+\xi_{4}, & \omega_2 +\kappa_2 -\xi_2 , & \xi_{3}, & \xi_2 , & \frac{i}{2}(\xi_{3}-\xi_{4}), & -\frac{i}{2}(\kappa_2 -\xi_2 )\\
\xi_1 , & \xi_{3}, & \omega_1 +\kappa_1 -\xi_1 , & -\xi_{3}+\xi_{4}, & \frac{i}{2}(\kappa_1 -\xi_1 ), & -\frac{i}{2}(\xi_{3}-\xi_{4})\\
\xi_{3}, & \xi_2 , & -\xi_{3}+\xi_{4}, & \omega_2 +\kappa_2 -\xi_2 , & -\frac{i}{2}(\xi_{3}-\xi_{4}), & \frac{i}{2}(\kappa_2 -\xi_2 )\\
\frac{i}{2}(\kappa_1 -\xi_1 ), & -\frac{i}{2}(\xi_{3}-\xi_{4}), & -\frac{i}{2}(\kappa_1 -\xi_1 ), & \frac{i}{2}(\xi_{3}-\xi_{4}), & \kappa_1 +\omega_1 , & \xi_{4}\\
-\frac{i}{2}(\xi_{3}-\xi_{4}), & \frac{i}{2}(\kappa_2 -\xi_2 ), & \frac{i}{2}(\xi_{3}-\xi_{4}), & -\frac{i}{2}(\kappa_2 -\xi_2 ), & \xi_{4}, & \kappa_2 +\omega_2
\end{array}\right]$}
\end{equation}

\textbf{CP-even, $Z_2 $ even, \& $Q_{em}=0$}: the basis is \{$h\chi^{+}$,
$\chi^0\chi^{+}$, $S_1^0S_1^{+}$, $S_1^0S_2^{+}$,
$S_2^0S_1^{+}$, $S_2^0S_2^{+}$, $Q_1^0S_1^{+}$,
$Q_1^0S_2^{+}$, $Q_2^0S_1^{+}$, $Q_2^0S_2^{+}$\};
and the matrix is
\begin{equation}
\resizebox{0.93\textwidth}{!}{$\left[
\begin{array}{cccccccccc}
\frac{1}{3}\lambda_{H}, & 0, & \frac{1}{2}(\kappa_1 +\xi_1 ), & \frac{1}{2}(\xi_{3}+\xi_{4}), & \frac{1}{2}(\xi_{3}+\xi_{4}), & \frac{1}{2}(\kappa_2 +\xi_2 ), & \frac{i}{2}(\kappa_1 -\xi_1 ), & -\frac{i}{2}(\xi_{3}-\xi_{4}), & -\frac{i}{2}(\xi_{3}-\xi_{4}), & \frac{i}{2}(\kappa_2 -\xi_2 )\\
0, & \frac{1}{3}\lambda_{H}, & -\frac{i}{2}(\kappa_1 -\xi_1 ), & \frac{i}{2}(\xi_{3}-\xi_{4}), & \frac{i}{2}(\xi_{3}-\xi_{4}), & -\frac{i}{2}(\kappa_2 -\xi_2 ), & \frac{1}{2}(\kappa_1 +\xi_1 ), & \frac{1}{2}(\xi_{3}+\xi_{4}), & \frac{1}{2}(\xi_{3}+\xi_{4}), & \frac{1}{2}(\kappa_2 +\xi_2 )\\
\frac{1}{2}(\kappa_1 +\xi_1 ), & \frac{i}{2}(\kappa_1 -\xi_1 ), & \frac{1}{3}\lambda_1 , & 0, & 0, & \frac{1}{2}\varrho_2 , & 0, & 0, & 0, & \frac{i}{2}\varrho_2 \\
\frac{1}{2}(\xi_{3}+\xi_{4}), & -\frac{i}{2}(\xi_{3}-\xi_{4}), & 0, & \varrho_1 , & \frac{1}{2}\varrho_2 , & 0, & 0, & 0, & -\frac{i}{2}\varrho_2 , & 0\\
\frac{1}{2}(\xi_{3}+\xi_{4}), & -\frac{i}{2}(\xi_{3}-\xi_{4}), & 0, & \frac{1}{2}\varrho_2 , & \varrho_1 , & 0, & 0, & -\frac{i}{2}\varrho_2 , & 0, & 0\\
\frac{1}{2}(\kappa_2 +\xi_2 ), & \frac{i}{2}(\kappa_2 -\xi_2 ), & \frac{1}{2}\varrho_2 , & 0, & 0, & \frac{1}{3}\lambda_2 , & \frac{i}{2}\varrho_2 , & 0, & 0, & 0\\
-\frac{i}{2}(\kappa_1 -\xi_1 ), & \frac{1}{2}(\kappa_1 +\xi_1 ), & 0, & 0, & 0, & -\frac{i}{2}\varrho_2 , & \frac{1}{3}\lambda_1 , & 0, & 0, & \frac{1}{2}\varrho_2 \\
\frac{i}{2}(\xi_{3}-\xi_{4}), & \frac{1}{2}(\xi_{3}+\xi_{4}), & 0, & 0, & \frac{i}{2}\varrho_2 , & 0, & 0, & \varrho_1 , & \frac{1}{2}\varrho_2 , & 0\\
\frac{i}{2}(\xi_{3}-\xi_{4}), & \frac{1}{2}(\xi_{3}+\xi_{4}), & 0, & \frac{i}{2}\varrho_2 , & 0, & 0, & 0, & \frac{1}{2}\varrho_2 , & \varrho_1 , & 0\\
-\frac{i}{2}(\kappa_2 -\xi_2 ), & \frac{1}{2}(\kappa_2 +\xi_2 ), & -\frac{i}{2}\varrho_2 , & 0, & 0, & 0, & \frac{1}{2}\varrho_2 , & 0, & 0, & \frac{1}{3}\lambda_2
\end{array}\right]$}
\end{equation}

\textbf{CP-even, $Z_2 $ even, \& $Q_{em}=0$}: the basis is \{$hS_1^{+}$,
$hS_2^{+}$, $\chi^0S_1^{+}$, $\chi^0S_2^{+}$, $S_1^0\chi^{+}$,
$S_2^0\chi^{+}$, $Q_1^0\chi^{+}$, $Q_2^0\chi^{+}$\};
and the matrix is
\begin{equation}
\resizebox{0.93\textwidth}{!}{$\left[
\begin{array}{cccccccc}
\omega_1 , & 0, & 0, & 0, & \frac{1}{2}(\kappa_1 +\xi_1 ), & \frac{1}{2}(\xi_{3}+\xi_{4}), & -\frac{i}{2}(\kappa_1 -\xi_1 ), & \frac{i}{2}(\xi_{3}-\xi_{4})\\
0, & \omega_2 , & 0, & 0, & \frac{1}{2}(\xi_{3}+\xi_{4}), & \frac{1}{2}(\kappa_2 +\xi_2 ), & \frac{i}{2}(\xi_{3}-\xi_{4}), & -\frac{i}{2}(\kappa_2 -\xi_2 )\\
0, & 0, & \omega_1 , & 0, & \frac{i}{2}(\kappa_1 -\xi_1 ), & -\frac{i}{2}(\xi_{3}-\xi_{4}), & \frac{1}{2}(\kappa_1 +\xi_1 ), & \frac{1}{2}(\xi_{3}+\xi_{4})\\
0, & 0, & 0, & \omega_2 , & -\frac{i}{2}(\xi_{3}-\xi_{4}), & \frac{i}{2}(\kappa_2 -\xi_2 ), & \frac{1}{2}(\xi_{3}+\xi_{4}), & \frac{1}{2}(\kappa_2 +\xi_2 )\\
\frac{1}{2}(\kappa_1 +\xi_1 ), & \frac{1}{2}(\xi_{3}+\xi_{4}), & -\frac{i}{2}(\kappa_1 -\xi_1 ), & \frac{i}{2}(\xi_{3}-\xi_{4}), & \omega_1 , & 0, & 0, & 0\\
\frac{1}{2}(\xi_{3}+\xi_{4}), & \frac{1}{2}(\kappa_2 +\xi_2 ), & \frac{i}{2}(\xi_{3}-\xi_{4}), & -\frac{i}{2}(\kappa_2 -\xi_2 ), & 0, & \omega_2 , & 0, & 0\\
\frac{i}{2}(\kappa_1 -\xi_1 ), & -\frac{i}{2}(\xi_{3}-\xi_{4}), & \frac{1}{2}(\kappa_1 +\xi_1 ), & \frac{1}{2}(\xi_{3}+\xi_{4}), & 0, & 0, & \omega_1 , & 0\\
-\frac{i}{2}(\xi_{3}-\xi_{4}), & \frac{i}{2}(\kappa_2 -\xi_2 ), & \frac{1}{2}(\xi_{3}+\xi_{4}), & \frac{1}{2}(\kappa_2 +\xi_2 ), & 0, & 0, & 0, & \omega_2
\end{array}\right]$}
\end{equation}

\end{document}